\documentclass[
  aps,prl,
  reprint,
  superscriptaddress,
  amsmath,amssymb,
  floatfix,
  nofootinbib
]{revtex4-2}

\usepackage{graphicx}
\usepackage{bm}
\usepackage{xcolor}
\usepackage{hyperref}
\hypersetup{colorlinks=true,
            linkcolor=blue!55!black,
            citecolor=blue!55!black,
            urlcolor=blue!55!black}
\renewcommand{\selectlanguage}[1]{}
\usepackage{amsthm}
\usepackage[capitalize,nameinlink]{cleveref}

\theoremstyle{plain}

\theoremstyle{definition}

\theoremstyle{remark}

\newcommand{\Dis}{\mathcal{D}}
\newcommand{\Lv}{\mathcal{L}}
\newcommand{\Tr}{\operatorname{Tr}}
\newcommand{\heff}{h_{\mathrm{eff}}}
\newcommand{\Leg}{\mathsf{P}}
\newcommand{\Pmin}{P_{\min}}

\begin{document}

\title{Optimal Scaling of Thermodynamic Uncertainty in Fermion Chains}

\author{Zhiyuan Yao}
\email{yaozy@lzu.edu.cn}
\affiliation{Lanzhou Center for Theoretical Physics, Key Laboratory of Theoretical Physics of Gansu Province, Key Laboratory of Quantum Theory and Applications of MoE, Gansu Provincial Research Center for Basic Disciplines of Quantum Physics, Lanzhou University, Lanzhou 730000, China}

\date{\today}

\begin{abstract}
In classical Markov dynamics, the thermodynamic uncertainty relation bounds
the product $P$ of entropy production and relative current variance from below
by $2$. Quantum coherence can violate this bound, yet how fast $P$ can fall
with system size and at what resource cost has remained open. For chains of
$n$ noninteracting fermionic modes, precision is limited by how sharply
transmission can switch between $0$ and $1$ in energy: we prove that $P$ falls
no faster than $\log n/n^2$, and construct chains attaining this optimal
scaling at every length with finite current. Moreover, the contact linewidth
per transferred particle must grow logarithmically with $1/P$, which is
achievable; for bounded contacts, $P$ can vanish only at vanishing current.
Coherently driven open quantum systems coupled to thermal baths therefore have
no positive lower bound on $P$: in Hilbert-space dimension $N$, the minimal
product decays at least as fast as $\log\log N/(\log N)^2$.
\end{abstract}

\maketitle

\textit{Introduction}.---Every device that runs out of equilibrium, from a
molecular motor to a thermoelectric junction or a clock, delivers a current
that fluctuates. In classical Markov dynamics the thermodynamic uncertainty
relation bounds the uncertainty product $P$ of any steady current from below
by $2$~\cite{Barato_2015tu,Gingrich_2016db,Horowitz_2019tu}. Precision thus
has a minimal entropy price, which constrains engines and
transport~\cite{Pietzonka_2018ut} and the accuracy of autonomous
clocks~\cite{Barato_2016ca,Erker_2017aq,Pearson_2021mt}.

Quantum coherence evades this price: coherent conductors and driven quantum
systems can produce currents more precise than any classical process at the
same dissipation~\cite{Ptaszynski_2018ce,Agarwalla_2018at,Brandner_2018tb,
Kalaee_2021vt,Rignon-Bret_2021to,Ehrlich_2021bf,Sundelin_2026aq}. What then
limits precision? Menczel, Loisa, Brandner and Flindt conjectured that for
coherently driven devices exchanging energy with thermal baths no floor on
$P$ survives as the device grows~\cite{Menczel_2021tu}. If so, the questions
become how fast $P$ can fall with the size of a device, and which physical
resources this consumes.

Energy filtering is the natural route~\cite{Ehrlich_2021bf,Timpanaro_2025qt,
Brandner_2025tb}. A conductor transmitting perfectly inside an energy window
and not at all outside it has no partition noise, and a large bias removes
thermal noise. Chains of quantum dots approach such a window. For the trigonometric chain of Ref.~\cite{Brandner_2025tu}, $P$ was shown numerically to vanish as the bias and the number of dots grow, though only about as fast as the inverse number of dots~\cite{Meier_2026eg} and with coupling ratios that grow with the length~\cite{Brandner_2025tu}; numerically optimized chains do better~\cite{Meier_2026eg}. Existing bounds do not decide how fast $P$ can fall. Size-dependent bounds are known when the environment is finite~\cite{Hasegawa_2025fp}, but not for devices driven by Markovian reservoirs, and bounds through the
minimal reflection~\cite{Palmqvist_2025ck} become empty at near-perfect
transmission. What such transmission demands of the contacts was not
known either~\cite{Gerry_2022aa}.

Here we determine both how fast $P$ can fall and what this consumes, for chains of noninteracting fermions. The transmission of an $n$-mode chain is $1/(1+Q)$ with $Q$ a polynomial of degree $2n$, whose slope Markov's inequality bounds, so partition noise survives at the edges of every transmission window. This fixes the size law $P_{\min}(n)=\Theta(\log n/n^2)$, attained by a chain with a Gegenbauer transmission profile. A sum rule for $\log[1/(1-T)]$, akin to Fano's matching limit for electrical networks, fixes the contact law: the contact linewidth per transferred particle must grow as $\tfrac12\log(1/P)$, as Chebyshev chains achieve, and bounded contacts give unlimited precision only at a vanishing current. Embedding the chains in larger Hilbert spaces proves the conjecture of Menczel et al.\ with the rate $\log\log N/(\log N)^2$. The size law is a noise version of a classical question of filter design, how sharply a filter of given order can switch~\cite{Cohn_1957dc,Stojanovic_2016dt}.

\textit{Thermal current through a chain}.---A precise current needs two
noise sources removed at once: the randomness of transmission and the reverse
jumps of the reservoirs. We first define the class of dynamics in which the question is posed, and then the chain in which both noise sources can be computed. The thermal class, the part of
the class of Ref.~\cite{Menczel_2021tu} with positive gaps and thermal
occupations, consists of Lindblad
generators~\cite{Lindblad_1976ot,Gorini_1976cp}, with $\hbar=k_B=1$ and $\log$ the natural logarithm,
\begin{equation}
\Lv\rho=-i[H_0+X,\rho]+\sum_\mu\gamma_\mu\big[(\bar n_\mu+1)\Dis[L_\mu]\rho
+\bar n_\mu\Dis[L_\mu^\dagger]\rho\big],
\label{eq:class}
\end{equation}
with $\Dis[L]\rho=L\rho L^\dagger-\tfrac12\{L^\dagger L,\rho\}$,
an arbitrary coherent drive $X$, and paired channels lowering the bare
energy, $[H_0,L_\mu]=-\Delta E_\mu L_\mu$, with Bose occupations
$\bar n_\mu=1/(e^{\beta_\mu\Delta E_\mu}-1)$ at positive finite inverse
temperatures $\beta_\mu$. Since this class lies inside that of Ref.~\cite{Menczel_2021tu}, an upper bound on the minimal $P$ proved in it holds there as well. Let $j_\mu$ be the mean rate of jumps $L_\mu$ minus that of jumps $L_\mu^\dagger$. The channel affinity is
$A_\mu=\beta_\mu\Delta E_\mu$, and the entropy
production rate is $\sigma=\sum_\mu A_\mu j_\mu$. Counting $L_\mu$ as $+1$ and
$L_\mu^\dagger$ as $-1$ in one channel gives a net count $N_t$ with mean rate $j$ and
variance rate $d=\lim_{t\to\infty}\mathrm{Var}(N_t)/t$, so that the classical bound reads $P\ge2$ for the uncertainty product
$P=\sigma d/j^2$. We write $P_{\min}^N$ for its infimum in Hilbert-space
dimension $N$, over members with a unique attracting stationary state and a
positive counted current.

Our device is a chain of $n$ fermionic modes $c_1,\dots,c_n$ with Hamiltonian
$H=\sum_{l=1}^{n-1}t_l(c_l^\dagger c_{l+1}+c_{l+1}^\dagger c_l)$, positive
hoppings $t_l$ and zero on-site energies. Particles enter at mode $1$ and
leave at mode $n$:
\begin{equation}
\begin{split}
\Lv\rho={}&-i[H,\rho]+2\Gamma\big\{f_L\Dis[c_1^\dagger]+f_R\Dis[c_1]\big\}\rho\\
&+2\Gamma\big\{f_L\Dis[c_n]+f_R\Dis[c_n^\dagger]\big\}\rho,
\end{split}
\label{eq:chain}
\end{equation}
with $f_L=1/(1+e^{-A})$, $f_R=1-f_L$, and contact linewidth $\Gamma$ at both
ends. We count extraction at mode $n$ positively. With $H_0=A(I-\hat n_1+\hat
n_n)$ and $X=H-H_0$, both forward jumps $c_1^\dagger$ and $c_n$ lower $H_0$
by $A$ at temperature $1$, so Eq.~\eqref{eq:chain} is a member of the class
\eqref{eq:class} in dimension $2^n$, with $A$ the affinity of both channels.
For every finite $A>0$ it has a unique stationary state, which is faithful,
and every density matrix relaxes to it~\cite{supplement}.

Let $h$ be the single-particle hopping matrix and
$T(E)=4\Gamma^2|(E-h+i\Gamma\Pi_1+i\Gamma\Pi_n)^{-1}_{n1}|^2$ the
transmission, with $\Pi_l=|l\rangle\langle l|$. Write
$I_1=\int T\,dE/2\pi$ and $I_2=\int T(1-T)\,dE/2\pi$, and let $F_0=I_2/I_1$
be the Fano factor at full bias. Then $j=qI_1$ with $q=\tanh(A/2)$, every net
transfer produces entropy $2A$, and
\begin{equation}
P=2Aq\,F_0+\frac{2A}{\sinh A}.
\label{eq:product}
\end{equation}
These finite-affinity formulas were obtained by Meier et al.\ in a
scattering treatment~\cite{Meier_2026eg}; we derive them from the full-Fock
generator~\eqref{eq:chain}, through the tilted generator and the Gaussian
stationary state~\cite{supplement}. The first term is partition noise, which
needs $F_0\to0$, a transmission close to $0$ or $1$ at almost every energy.
The second is thermal noise from reverse jumps. It needs $A\to\infty$, but
then the first term grows as $A$. Both must vanish together.

The trigonometric chain of Ref.~\cite{Brandner_2025tu} shows the balance.
Its transmission is $1/(1+E^{2n})$ at every length~\cite{supplement}, so $F_0=1/(2n)$. At $A=\log n$ the two
terms of Eq.~\eqref{eq:product} are $(\log n)/n$ and $4(\log n)/n$
asymptotically, so $P\simeq5\log n/n$. The entropy per transferred particle
grows as $2\log n$, while the Fano factor $d/j$ falls as $5/(2n)$. The rate
of decrease is thus set by how fast $F_0$ can fall with $n$.

\textit{How sharp can a chain filter be?}---Partition noise lives where
transmission is neither near zero nor near one, so it is set by how steeply an
$n$-mode chain can switch transmission on and off. Here we allow any real
on-site energies and any linewidths $\Gamma_L,\Gamma_R$ of the two end
contacts; Eq.~\eqref{eq:product} still holds~\cite{supplement}. The
end-to-end cofactor of $E-h+i\Gamma_L\Pi_1+i\Gamma_R\Pi_n$ is the constant
$\prod_lt_l$, whatever the on-site energies and linewidths. Unitarity of the
two-terminal scattering then gives
\begin{equation}
\begin{gathered}
T(E)=\frac{1}{1+Q(E)},\\
Q(E)=\frac{|\det(E-h-i\Gamma_L\Pi_1+i\Gamma_R\Pi_n)|^2}{4\Gamma_L\Gamma_R\prod_lt_l^2},
\end{gathered}
\label{eq:Q}
\end{equation}
where $Q\ge0$ is a real polynomial of degree exactly $2n$, for every chain
with nonzero hoppings~\cite{supplement}. Transmission is near one where $Q$
is small and near zero where $Q$ is large, and the noise
$T(1-T)=Q/(1+Q)^2$ is at least $2/9$ wherever $\tfrac12\le Q\le1$.

Take an interval of length $\ell$ on which $0\le Q\le1$, with $Q=1$ at its
ends. Markov's inequality bounds the derivative of a polynomial of degree
$m$ on an interval by $m^2$ times its maximum, divided by half the
length~\cite{Markov_1890oa,Aptekarev_2014ao}. Applied to $2Q-1$, it gives
$|Q'|\le m^2/\ell$. Hence $Q$ stays above $\tfrac12$ within $\ell/(2m^2)$ of
each end, and these strips carry noise at least $2\ell/(9m^2)$, against a
transmitted weight at most $\ell$. Outside $\{Q\le1\}$, $T(1-T)\ge T/2$; if
$Q>1$ everywhere, this holds at every energy. Summing over the intervals of
$\{Q\le1\}$ and the rest of the axis, with $m=2n$,
\begin{equation}
F_0\ge\frac{1}{18n^2}.
\label{eq:F0lower}
\end{equation}
The bound needs no symmetry and holds at any coupling ratio and current.
It does not depend on the energy scale either: rescaling $h$, $\Gamma_L$ and $\Gamma_R$
together rescales the window and leaves $F_0$ unchanged. A chain has $n$
resonances to shape its transmission, and each window edge costs a fraction
$1/n^2$ of the window, however the resonances are placed.
Inserting Eq.~\eqref{eq:F0lower} in Eq.~\eqref{eq:product} and splitting at $A=\log n$ gives
$P\ge\log n/(18n^2)$ for $n\ge3$ and every finite $A$: above $\log n$ the
partition term exceeds it, and below, the thermal term
$2A/\sinh A\ge4n\log n/(n^2-1)$ does~\cite{supplement}.

\begin{figure}
\includegraphics[width=\columnwidth]{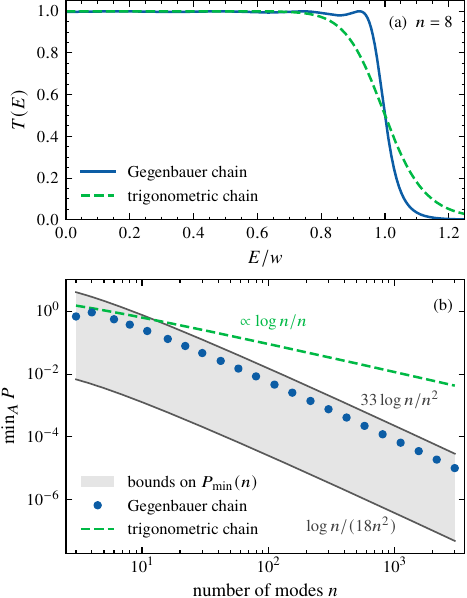}
\caption{Size law. (a) Transmission of the Gegenbauer chain,
Eq.~\eqref{eq:gegenbauer}, and of the trigonometric
chain~\cite{Brandner_2025tu} with $n=8$ modes. The ripple of the Gegenbauer
chain is small and largest next to the window edge. Its edge slope is within a factor
$4$ of the Markov limit. (b) Uncertainty product minimized over the affinity, for
both chains, against the number of modes. The shaded band lies between the
bounds~\eqref{eq:sizelaw} on $P_{\min}(n)$; the Gegenbauer chain falls as
$\log n/n^2$ inside it, while a single chain may lie above it, as the trigonometric chain does, falling as $\log n/n$. Products are
computed from Eq.~\eqref{eq:product} with $F_0$ integrated over the whole
energy axis.}
\label{fig:size}
\end{figure}

\textit{A filter at the limit}.---The bound~\eqref{eq:F0lower} asks for a
profile whose small values are spread over the window and whose rise
to $1$ is close to as steep as Markov's inequality permits. Let $\mathsf P_k$ be the
Legendre polynomial with $\mathsf P_k(1)=1$, and
\begin{equation}
\begin{gathered}
r_n(x)=\frac{2\mathsf P_{n+1}'(x)}{(n+1)(n+2)}
=\frac{C^{(3/2)}_n(x)}{C^{(3/2)}_n(1)},\\
T_n(E)=\frac{1}{1+r_n(E/w)^2},
\end{gathered}
\label{eq:gegenbauer}
\end{equation}
a normalized Gegenbauer polynomial, with $w$ the half-width of the window.
The same normalized Gegenbauer prototype appears in filter
synthesis~\cite{Stojanovic_2016dt}. Orthogonality gives
$\int_{-1}^1r_n^2\,dx=4/[(n+1)(n+2)]$, so the in-band ripple carries little
weight; $|r_n|\le1$ on the window, with its maximum $1$ at the edges $x=\pm1$. Beyond the
edge every coefficient of the expansion of $r_n(1+u)$ in $u$ is positive,
so $r_n(1+u)\ge1+n(n+3)u/4$. The slope of $Q=r_n^2$ at the edge is thus
within a factor $4$ of the Markov limit. Together these give
\begin{equation}
F_0\le\frac{6}{n(n+3)}+\frac{8}{n^2(n+3)^2}<\frac{6}{n^2},
\label{eq:F0upper}
\end{equation}
including both infinite tails~\cite{supplement}. Figure~\ref{fig:size}(a)
compares $T_8$ with the transmission $1/(1+E^{2n})$ of the trigonometric
chain~\cite{Brandner_2025tu}, whose edges are wider by a factor of order
$n$ and whose $F_0=1/(2n)$. The chains optimized numerically in
Ref.~\cite{Meier_2026eg}, whose full-bias Fano factor falls about as
$n^{-1.86}$, are consistent with Eq.~\eqref{eq:F0lower}, which forbids any
exponent above $2$.

A transmission $T_n$ is useful only if a chain of the form~\eqref{eq:chain}
produces it. For every $n\ge2$ and $w>0$ it does. Write $r_n(E/w)=R(E)/\kappa$
with $R$ monic, and let $D$ be the monic polynomial whose roots are the
roots of $R^2+\kappa^2$ in the lower half-plane. Write $D=U+iV$ with $U$ and
$V$ real polynomials. Then $\Gamma$ is half the
coefficient of $E^{n-1}$ in $V$. The polynomial $(U+R)/2$ has $n$ simple real
roots $x_k$, which carry positive weights $\kappa/[2\Gamma|(U+R)'(x_k)/2|]$
summing to one. The Jacobi matrix of this discrete
measure~\cite{Arlinskii_2006ns} has zero diagonal, positive mirror-symmetric
hoppings, and transmission exactly $T_n(E)$ at every real
$E$~\cite{supplement}.

Choosing $A=2\log n$ in Eq.~\eqref{eq:product} makes the thermal term
$8\log n/[n^2(1-n^{-4})]$. Let $P_{\min}(n)$ be the infimum of $P$ over all
chains of $n$ fermionic modes and all finite $A>0$. The chains have nonzero
real nearest-neighbor hoppings, real on-site energies, and Markovian
contacts of linewidths $\Gamma_L,\Gamma_R>0$ on the end modes. Inserting Eqs.~\eqref{eq:F0lower} and \eqref{eq:F0upper} in
Eq.~\eqref{eq:product} gives the size law: for every $n\ge3$,
\begin{equation}
\frac{\log n}{18n^2}\le P_{\min}(n)<33\,\frac{\log n}{n^2}.
\label{eq:sizelaw}
\end{equation}
The upper bound is realized, for every $n\ge2$ and $w>0$, by the chain with
transmission~\eqref{eq:gegenbauer} at $A=2\log n$, whose current satisfies
$|\pi j/w-1|\le4/n^2$~\cite{supplement}.

Thus the product falls as $\Theta(\log n/n^2)$, and the order-optimal chain keeps a finite current: at fixed $w$ it tends to $w/\pi$. Figure~\ref{fig:size}(b) shows the product
minimized over $A$ inside the proven band; the trigonometric chain falls only
as $\log n/n$. The remaining factor between the constants $1/18$ and $33$ is
not resolved. Equations~\eqref{eq:F0lower} and~\eqref{eq:F0upper} also show that the lower bound on $F_0$ is sharp up to a constant factor.

\textit{What the contacts must supply}.---Precision needs near-perfect
transmission over a window of energies. A contact of finite linewidth cannot
match over a broad window, a limit Fano found for electrical
networks~\cite{Fano_1950tl}. We state it for quadratic networks
with two single-mode contacts, into which every mode decays. The single-particle
Hamiltonian is an $n\times n$ Hermitian $h$,
with on-site energies and complex hoppings allowed. Contacts of linewidths
$\Gamma_L$ and $\Gamma_R$ act on modes $1$ and $n$. We require
$-ih-\Gamma_L\Pi_1-\Gamma_R\Pi_n$ to be strictly stable and the transmission
to be nonzero. The generator
then has a unique attracting stationary state and obeys
Eq.~\eqref{eq:product}~\cite{supplement}. Let $z_k$ be the eigenvalues of
$h+i\Gamma_L\Pi_1-i\Gamma_R\Pi_n$, the zeros of the reflection determinant
counted with any factors it shares with the denominator. The poles and these
zeros give
\begin{equation}
\int\!\frac{dE}{2\pi}\log\frac{1}{1-T(E)}
=\Gamma_L+\Gamma_R-\sum_k|\operatorname{Im}z_k|
\le2\Gamma_*,
\label{eq:budget}
\end{equation}
with $\Gamma_*=\min(\Gamma_L,\Gamma_R)$, because
$\sum_k|\operatorname{Im}z_k|\ge|\Gamma_L-\Gamma_R|$. A fixed budget of
$\log[1/(1-T)]$ is thus available. Spending it on high transmission over a
window narrows the window: a window of width $\Delta E$ with $1-T\le\epsilon$
costs $\Delta E\log(1/\epsilon)/2\pi$. Partition noise is the average of
$1-T$ over the transmitted weight, $F_0=\int T(1-T)/\int T$. Jensen's
inequality for the convex function $\log(1/x)/(1-x)$, applied with that
weight, turns Eq.~\eqref{eq:budget} into~\cite{supplement}
\begin{equation}
I_1\,\frac{\log(1/F_0)}{1-F_0}\le2\Gamma_*.
\label{eq:jensen}
\end{equation}
A small $F_0$ at a given current thus requires a contact
linewidth that grows as $\log(1/F_0)$. The bound is set by the weaker
contact. Here $\Gamma_*$ is the linewidth that contact gives the end mode,
not a rate of reservoir jumps. Unlike bounds through the minimal
reflection~\cite{Palmqvist_2025ck}, Eq.~\eqref{eq:jensen} stays informative
when transmission is near perfect, and it gives the contact linewidth that near-perfect transmission requires; without such transmission a cost-precision trade-off returns~\cite{Gerry_2022aa}.

Combined with Eq.~\eqref{eq:product}, this budget bounds $\Gamma_*/j$ from
below at a given product $p$. Write $s=\log(1/p)$, and let $\mathcal R(p)$ be the
minimum of $\log(1/f)/[2q(1-f)]$ with $f=(p-2A/\sinh A)/(2Aq)$, over the
affinities with $2A/\sinh A<p$. The
thermal term $2A/\sinh A>e^{-A}$ forces $A>s$. Partition noise is then
allowed only $F_0\le p/(2Aq)$, smaller than $p$ by a factor at least of order $s$, and
Eq.~\eqref{eq:jensen} turns this factor into the second
logarithm. For $0<p<2$, every two-contact network as above with $P\le p$
thus obeys
\begin{equation}
\Gamma_*/j\ge\mathcal R(p),
\label{eq:contactR}
\end{equation}
and, for $s=\log(1/p)\ge8$~\cite{supplement},
\begin{equation}
0\le2\mathcal R(p)-\big[s+\log(2s)\big]\le\frac{2\log s+\log8+5}{s}.
\label{eq:contactbound}
\end{equation}
The bound is sharp up to an additive constant: for every
$0<p\le1$ an equal-contact chain with $P<p$ exists, and as $p\to0$ its cost is
$\Gamma/j=s/2+\tfrac12\log s+\log8+o(1)$. The infimum of the contact linewidth per
transferred particle over two-contact networks of any size, at any finite
affinity, with $P\le p$ is therefore, as $p\to0$,
\begin{equation}
\inf\frac{\Gamma_*}{j}=\frac12\log\frac1p+\frac12\log\log\frac1p+O(1),
\label{eq:contactlaw}
\end{equation}
the contact law.

\begin{figure}
\includegraphics[width=\columnwidth]{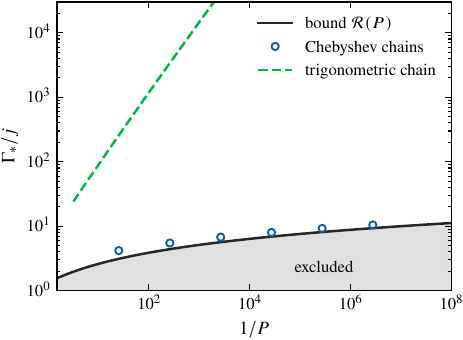}
\caption{Contact law. Weaker contact linewidth per transferred particle
needed for uncertainty product $P$. The shaded region below
$\mathcal R(P)$ is excluded for every two-contact network
[Eq.~\eqref{eq:contactR}]. Open circles: Chebyshev chains built for
$p=10^{-1},\dots,10^{-6}$, plotted at their actual product. Dashed: the
trigonometric chain, optimized over length and affinity.}
\label{fig:contact}
\end{figure}

The upper side of this law is realized by the Chebyshev transmission
$T(E)=[1+\epsilon^2\mathsf T_n(E/w)^2]^{-1}$, with $\mathsf T_n$ the Chebyshev polynomial of
the first kind. Its order $n$ is odd and of order $\log(1/p)/\sqrt p$, and its
ripple is $\epsilon^2=p/(16A)$ at the affinity $A=\log(16/p)+\log\log(16/p)$. Its hoppings are the classical prototype of
coupled-resonator filters~\cite{Cohn_1957dc}; we prove that they give this
transmission exactly with equal Markovian contacts~\cite{supplement}.
The trigonometric chain, optimized over length and affinity, needs
$\Gamma/j=[s+2\log s+\log4+1+o(1)]/p$, exponentially more
(Fig.~\ref{fig:contact}).

\textit{No dimension-independent precision floor}.---Menczel et al.\ asked for the
minimal product at a fixed Hilbert-space dimension $N$, over the whole thermal
class~\cite{Menczel_2021tu}. The Gegenbauer chain that realizes the upper bound of Eq.~\eqref{eq:sizelaw} has dimension
$2^n$. Attaching auxiliary levels by thermal channels of small occupation, as in
Ref.~\cite{Menczel_2021tu}, reaches all remaining dimensions at no extra
entropy production and a current above $w/(4\pi)$~\cite{supplement}. For every integer
$N\ge4$, with $n=\lfloor\log_2N\rfloor$,
\begin{equation}
0\le P_{\min}^N<66\,\frac{\log n}{n^2},\qquad
P_{\min}^N=O\!\left(\frac{\log\log N}{(\log N)^2}\right).
\label{eq:dimension}
\end{equation}
This proves the conjecture of Ref.~\cite{Menczel_2021tu} that
$P_{\min}^N\to0$. This rate is an upper bound over the whole thermal class.
The size law~\eqref{eq:sizelaw} is proved only for chains, and no lower bound
in $N$ is known for the class.

\textit{Precision at bounded couplings}.---The two laws price precision in
different resources. In the Gegenbauer chain at $A=2\log n$ the entropy per
transferred particle grows as $4\log n$ while the relative noise falls as
$1/n^2$. In the Chebyshev chain the hoppings in the middle stay near $w/2$, while
$\Gamma$ grows as $w\log(1/P)/(2\pi)$, so its contact-to-bulk ratio grows
with the precision. Can precision grow with every coupling bounded? Equation~\eqref{eq:jensen} with $j\le I_1$ gives
$F_0\ge e^{-2\Gamma_*/j}$, so at every affinity
\begin{equation}
P\ge2\tanh\tfrac12\,e^{-2\Gamma_*/j}.
\label{eq:floor}
\end{equation}
A finite current through bounded contacts keeps a positive floor: along
every sequence of two-contact networks with bounded $\Gamma_*$ and $P\to0$,
the current tends to zero. With a current tending to zero, the floor tends to
zero, and suitable chains do drive the product to zero. There are chains with
zero on-site energies, every hopping in $[1,6]$, both linewidths equal to
$1$, and lengths $m_k\to\infty$, whose products at finite affinities tend to
zero~\cite{supplement}.

Bounded couplings can make an arbitrarily sharp filter at the price of
bandwidth. A strong bond $v$ followed by a weak bond $1$ passes an amplitude
$1/v$ at the band center, so $\ell$ such pairs act as one hopping $v^{-\ell}$. Such hoppings carry transfer across dimerized chains~\cite{Zurita_2023fq,Palavicini_2026hq}. Likewise, $\nu$ pairs of
bonds $(3,1)$ between a contact and the chain reduce the unit linewidth to
$9^{-\nu}$. With $v\le6$, such blocks approximate, as their depth grows, any chain with zero on-site energies and equal contacts, scaled down in energy. Here the target is a Chebyshev chain of growing order. Every energy is scaled by $\delta_\nu\propto9^{-\nu}$, and the current falls by the same factor. The error must be small on the whole energy axis against the
shrinking transmitted weight, and narrow resonances outside the window
could carry comparable weight. At such resonances either the dressed contact or the effective coupling across the central blocks is exponentially small. In either case the transmission there is at most an exponentially small factor times the spectral density of the end sites, whose integral is fixed~\cite{supplement}.

In this construction the current at a fixed target order $n$ falls exponentially with the length, at a rate of order $1/n$. Equation~\eqref{eq:floor} requires only $j\le2\Gamma_*/\log(2\tanh\tfrac12/P)$. Along the whole sequence no rate in the length is known.

\textit{Conclusions}.---Quantum coherence removes the size-independent floor on the uncertainty product, but precision is not free: in chains of coherent modes the entropy per transferred particle still grows logarithmically in the length while the relative noise falls as its inverse square, and the contacts must supply a linewidth per transferred particle that grows as the logarithm of the precision, or else the current must vanish. The precision a chain of given length can reach is set by how sharply a filter of given order can switch. This bound is testable in quantum-dot arrays, already proposed as broadband energy filters~\cite{Ehrlich_2021bf}: their Fano factor at full bias cannot fall faster than $1/n^2$ in the number $n$ of dots, a prediction that, to our knowledge, has not been tested. The same accounting may apply
to coherent clocks, whose ticks are counted currents. Three questions remain open: whether one device can meet both prices at once, how fast precision can grow with length at bounded couplings, and how fast it can grow beyond noninteracting chains, as in the ring clock of Ref.~\cite{Meier_2025pi}.

\textit{Acknowledgments}.---Z.\ Y.\ acknowledges support from the Strategic
Priority Research Program of the Chinese Academy of Sciences under Grant
No.~XDB1680102. Z.\ Y.\ also acknowledges support by National Natural Science
Foundation of China (Grant No.~12247101). The author also acknowledges Gewu SciLab for providing the collaborative research environment in
which this project was developed.

OpenAI GPT-5.6 Sol, OpenAI GPT-6 Astra, Anthropic Claude Opus~5.5, and
Anthropic Claude Fable~5.1 were used throughout this work. They assisted in
developing and checking arguments, reviewing proofs and their hypotheses,
and organizing and typesetting the manuscript in LaTeX. The author takes
full responsibility for the content of this paper.

\textit{Data availability}.---The data and the code that support the
findings of this article are available from the author upon reasonable
request.

\clearpage
\onecolumngrid
\setcounter{secnumdepth}{3}
\setcounter{equation}{0}
\setcounter{figure}{0}
\setcounter{table}{0}
\setcounter{section}{0}
\setcounter{theorem}{0}
\renewcommand{\theequation}{S\arabic{equation}}
\renewcommand{\thefigure}{S\arabic{figure}}
\renewcommand{\thetable}{S\arabic{table}}
\renewcommand{\thesection}{\Roman{section}}
\renewcommand{\thetheorem}{S\arabic{theorem}}
\renewcommand{\theHequation}{S\arabic{equation}}
\renewcommand{\theHfigure}{S\arabic{figure}}
\renewcommand{\theHtable}{S\arabic{table}}
\renewcommand{\theHsection}{S\Roman{section}}
\renewcommand{\theHtheorem}{S\arabic{theorem}}

\begin{center}
{\large\bfseries Supplemental Material for ``Optimal Scaling of Thermodynamic Uncertainty in Fermion Chains''}
\end{center}
\vspace{1em}

This Supplemental Material contains the proofs of the results stated in the
main text. Sections~\ref{sec:nonneg}--\ref{sec:cumulants} establish, for every
finite chain and every strictly stable two-contact network at finite affinity,
the nonnegativity of the product, a unique faithful attracting stationary
state on the full Fock space, and the exact current and variance.
Section~\ref{sec:lower} writes the transmission of a chain as $1/(1+Q)$ and
derives the lower bound on partition noise from Markov's inequality.
Sections~\ref{sec:gegenbauer}--\ref{sec:sizelaw} construct the Gegenbauer chain
at every length and assemble the size law. Sections~\ref{sec:budget} and
\ref{sec:contactcost} derive the contact budget of two-contact networks and the
minimal contact cost, which an equal-contact Chebyshev chain attains. Section~\ref{sec:bounded} shows that the product can vanish with every coupling bounded, and that it then forces the current to vanish.
Section~\ref{sec:padding} extends the upper bound of the size law to every Hilbert-space dimension. Section~\ref{sec:transmission} computes the transmission of the
trigonometric chain used as a benchmark, and Sec.~\ref{sec:calibration}
calibrates the counting conventions against the small-system values of Menczel
et al.~\cite{Menczel_2021tu}.

We follow the notation of the main text. The number of modes is $n\ge2$,
$f_{\rm L}=1/(1+e^{-A})$, $f_{\rm R}=1-f_{\rm L}$, $q=\tanh(A/2)=f_{\rm
L}-f_{\rm R}$, and $\Pi_j=|j\rangle\langle j|$ is a projector of the
single-particle space $\mathbb C^n$. We write $\beta(A)=2A/\sinh A$ for the
thermal term of the product, and $\log$ is the natural logarithm. The Legendre
polynomial of degree $k$ is written $\Leg_k$, normalized by $\Leg_k(1)=1$, to
keep it apart from the product $P$; $\mathsf T_k$ is the Chebyshev polynomial of the
first kind, $\mathsf T_k(\cos z)=\cos(kz)$.

\section{Nonnegativity of the product}
\label{sec:nonneg}

The bounds of the main text are bounds on a product that must first be shown
never to be negative. This fixes the lower end of the dimension bound of the
main text. The class is that of the main text: Lindblad generators with bare
Hamiltonian $H_0$, drive $X$, and paired channels $L_\mu$, $L_\mu^\dagger$
with rates $\gamma_\mu(n_\mu+1)$, $\gamma_\mu n_\mu$ and affinities
$A_\mu=\log[(n_\mu+1)/n_\mu]$. Every member of the class produces entropy at a
nonnegative rate, so the product of every counted channel with $j_\mu\ne0$ is
nonnegative, and so is its infimum in every dimension:
\begin{equation}
\sigma\ge0,\qquad P_\mu\ge0,\qquad \Pmin^N\ge0 .
\label{eq:nonneg}
\end{equation}

To see this, let $\rho$ be the stationary state and $v$ a vector with
$\rho v=0$. The diagonal element of $\Lv\rho=0$ at $v$ is
$0=\sum_\mu\gamma_\mu[(n_\mu+1)\langle v,L_\mu\rho L_\mu^\dagger v\rangle+n_\mu\langle v,L_\mu^\dagger\rho L_\mu v\rangle]$,
because the Hamiltonian and anticommutator terms vanish at $v$. Every term is
nonnegative, so $\rho^{1/2}L_\mu^\dagger v=\rho^{1/2}L_\mu v=0$. The kernel of
$\rho$ is therefore invariant under every $L_\mu$ and $L_\mu^\dagger$, and so
is the support of $\rho$. On the support write
$\rho=\sum_ip_i|i\rangle\langle i|$ with $p_i>0$, and put
$w^\mu_{ij}=|\langle i|L_\mu|j\rangle|^2$, $\alpha_\mu=n_\mu+1$,
$\alpha'_\mu=n_\mu$. Inserting the dissipators into $-\Tr(\dot\rho\log\rho)$
channel by channel gives the entropy flows and currents
\begin{equation}
\begin{gathered}
\dot{\mathcal S}_\mu=\gamma_\mu\sum_{ij}w^\mu_{ij}(\alpha_\mu p_j-\alpha'_\mu p_i)\log\frac{p_j}{p_i},\\
j_\mu=\gamma_\mu\sum_{ij}w^\mu_{ij}(\alpha_\mu p_j-\alpha'_\mu p_i),
\end{gathered}
\end{equation}
with sums over the support. The Hamiltonian contributes nothing, since
$\Tr([H,\rho]\log\rho)=0$, and stationarity gives $\sum_\mu\dot{\mathcal S}_\mu=0$.
Adding this zero to $\sigma=\sum_\mu A_\mu j_\mu$ and using
$A_\mu=\log(\alpha_\mu/\alpha'_\mu)$,
\begin{equation}
\sigma=\sum_{\mu,ij}\gamma_\mu w^\mu_{ij}(\alpha_\mu p_j-\alpha'_\mu p_i)\log\frac{\alpha_\mu p_j}{\alpha'_\mu p_i}\ge0,
\end{equation}
because $(x-y)\log(x/y)\ge0$ for $x,y>0$. Each channel thus contributes a
relative-entropy-like sum of nonnegative terms. The variance rate is
nonnegative by definition, so $P_\mu\ge0$, which is Eq.~\eqref{eq:nonneg}.

\section{Unique stationary state and relaxation}
\label{sec:relaxation}

The current and its variance are long-time properties of steady operation,
so they are defined only if the device forgets how it was prepared. This
section shows that every chain relaxes to one stationary state from every
initial state, and so does every two-contact network whose single-particle
modes all decay. Coherences between Fock sectors of opposite parity are
included.

Let $h\in\mathbb C^{n\times n}$ be Hermitian, $H=\sum_{ab}h_{ab}c_a^\dagger c_b$
on the full $2^n$-dimensional Fock space, $\Gamma_{\rm L},\Gamma_{\rm R}>0$ and
$0<A<\infty$. The two-contact generator is
\begin{equation}
\Lv\rho=-i[H,\rho]+2\Gamma_{\rm L}\big(f_{\rm L}\Dis[c_1^\dagger]+f_{\rm R}\Dis[c_1]\big)\rho
+2\Gamma_{\rm R}\big(f_{\rm L}\Dis[c_n]+f_{\rm R}\Dis[c_n^\dagger]\big)\rho .
\label{eq:gen}
\end{equation}
Mode $1$ is filled at rate $2\Gamma_{\rm L}f_{\rm L}$ and emptied at rate
$2\Gamma_{\rm L}f_{\rm R}$; mode $n$ is emptied at rate $2\Gamma_{\rm R}f_{\rm
L}$ and filled at rate $2\Gamma_{\rm R}f_{\rm R}$, so every rate is positive at
finite $A$. The chain of the main text is the case of real tridiagonal $h$ with
zero diagonal, $h_{j,j+1}=h_{j+1,j}=t_j>0$, and
$\Gamma_{\rm L}=\Gamma_{\rm R}=\Gamma$. Networks allow complex hoppings,
arbitrary real on-site energies and unequal contacts. Put
\begin{equation}
\Sigma=\Gamma_{\rm L}\Pi_1+\Gamma_{\rm R}\Pi_n,\qquad K=-ih-\Sigma .
\label{eq:drift}
\end{equation}
The matrix $K$ is the single-particle drift: as shown in Sec.~\ref{sec:cumulants},
it generates the decay of the two-point functions. We call the network
strictly stable when every eigenvalue of $K$ has negative real part.

\subsection{Every mode of a chain decays}
\label{sec:stable}

A single-particle mode fails to decay only if it never reaches a contact.
Precisely, (i) $K$ is strictly stable if and only if no eigenvector of $h$
has vanishing first and last components. Hence (ii) every chain with nonzero
hoppings, real on-site energies and $\Gamma_{\rm L},\Gamma_{\rm R}>0$ is
strictly stable.

For (i), if $Kx=\lambda x$ with $x\ne0$, then
\begin{equation}
\operatorname{Re}\lambda\,\|x\|^2=\operatorname{Re}\langle x,Kx\rangle=-\Gamma_{\rm L}|x_1|^2-\Gamma_{\rm R}|x_n|^2\le0,
\label{eq:Relambda}
\end{equation}
since $\langle x,-ihx\rangle$ is imaginary. The real part therefore vanishes
only when $x_1=x_n=0$. Then $\Sigma x=0$ and $hx=i\lambda x$, so $x$ is an
eigenvector of $h$ invisible to both contacts. Conversely such an eigenvector
of $h$ with eigenvalue $\epsilon$ satisfies $Kx=-i\epsilon x$. For (ii), in a
chain an eigenvector with $x_1=0$ has, from the first row of $hx=\epsilon x$,
$t_1x_2=(\epsilon-\varepsilon_1)x_1=0$, and row by row $x_3=\dots=x_n=0$.
No eigenvector of $h$ therefore vanishes at the first mode.

Graph connectivity does not imply strict stability. On the diamond with
$h_{12}=h_{13}=h_{24}=h_{34}=1$, the vector $(0,1,-1,0)$ is an eigenvector of
$h$ with eigenvalue zero that vanishes at both contacts. Strict stability does
not imply transmission either. With $h_{12}=h_{13}=h_{24}=1$, $h_{34}=-1$, the
subspaces spanned by $\{e_1,(e_2+e_3)/\sqrt2\}$ and $\{e_4,(e_2-e_3)/\sqrt2\}$
are invariant, each is a damped dimer, and $G_{41}\equiv0$. The two hypotheses,
strict stability and $I_1>0$, are therefore stated separately below.

\subsection{A unique attracting stationary state on the full Fock space}
\label{sec:unique}

Decaying modes are enough for the whole many-body state to relax. Let $h$ be
Hermitian, $\Gamma_{\rm L},\Gamma_{\rm R}>0$, $0<A<\infty$, and let $K$ be
strictly stable. Then the generator~\eqref{eq:gen} has a unique stationary
state $\rho_*$, which is faithful; zero is an algebraically simple eigenvalue
of $\Lv$ on all $2^n\times2^n$ matrices, every other eigenvalue has negative
real part, and every matrix $\rho$ relaxes as
\begin{equation}
e^{t\Lv}\rho\to\rho_*\Tr\rho .
\label{eq:relax}
\end{equation}
By Sec.~\ref{sec:stable}, this holds for every chain with nonzero real
hoppings, real on-site energies and $\Gamma_{\rm L},\Gamma_{\rm R}>0$. The
argument has four steps. Only multiples of the identity commute with the
Hamiltonian and all jumps, so every stationary state is faithful. The
dissipation identity then excludes undamped oscillations, and contractivity
excludes growth.

Write the four jumps as $B\in\{c_1^\dagger,c_1,c_n,c_n^\dagger\}$ with rates
$w_B\in\{2\Gamma_{\rm L}f_{\rm L},2\Gamma_{\rm L}f_{\rm R},2\Gamma_{\rm R}f_{\rm
L},2\Gamma_{\rm R}f_{\rm R}\}$, all positive. For $x\in\mathbb C^n$ put
$c(x)=\sum_ax_ac_a$ and $c^\dagger(x)=\sum_ax_ac_a^\dagger$. The
anticommutation relations give
\begin{equation}
[H,c(x)]=-c(h^{\mathsf T}x),\qquad [H,c^\dagger(x)]=c^\dagger(hx).
\label{eq:Hc}
\end{equation}
The commutator with $H$ thus moves a contact mode through the network, along
$h^{\mathsf T}$ for annihilators and along $h$ for creators.

\textit{Step 1: the commutant is trivial.} Let $Z$ commute with $H$ and with the
four jumps. The set $V=\{x:[Z,c(x)]=0\}$ is a subspace containing $e_1$ and
$e_n$. By the Jacobi identity and Eq.~\eqref{eq:Hc},
$[Z,c(h^{\mathsf T}x)]=-[Z,[H,c(x)]]=0$ for $x\in V$, so $V$ is invariant under
$h^{\mathsf T}$. Its orthogonal complement is then invariant under
$(h^{\mathsf T})^\dagger=\bar h$ and orthogonal to $e_1$ and $e_n$. If it were
nonzero, it would contain an eigenvector $y$ of $\bar h$ with $y_1=y_n=0$, and
$\bar y$ would be an eigenvector of $h$ with the same real eigenvalue and the
same vanishing components, contradicting Sec.~\ref{sec:stable}. Hence
$V=\mathbb C^n$. The same argument with $V'=\{x:[Z,c^\dagger(x)]=0\}$, which is
invariant under $h$, gives $V'=\mathbb C^n$. So $Z$ commutes with every $c_a$
and $c_a^\dagger$. These generate the full matrix algebra: products of
$\hat n_a$ and $1-\hat n_a$ give the projections onto Fock basis states, and
single $c_a$ or $c_a^\dagger$ flip one occupation. So $Z$ is a multiple of the
identity.

\textit{Step 2: every stationary state is faithful.} A stationary density
matrix exists, as a limit point of the time averages
$\tau^{-1}\int_0^\tau e^{t\Lv}\rho\,dt$. Let $\rho$ be one, with support
projection $\mathsf P$. As in Sec.~\ref{sec:nonneg}, the diagonal elements of
$\Lv\rho=0$ on the kernel of $\rho$ force $(I-\mathsf P)B\mathsf P=0$ for all
four jumps. The set of jumps is closed under adjoints, so $[\mathsf P,B]=0$.
The dissipators then have no off-diagonal blocks with respect to $\mathsf P$,
and the off-diagonal block of $\Lv\rho=0$ reads $(I-\mathsf P)H\mathsf P\rho=0$.
As $\rho$ is invertible on its support, $(I-\mathsf P)H\mathsf P=0$, so
$[\mathsf P,H]=0$. By Step 1, $\mathsf P=I$.

\textit{Step 3: no other eigenvalue on the imaginary axis.} For any matrix
$Y$, expanding the adjoint generator $\Lv^\dagger$ gives the dissipation
identity
\begin{equation}
\Lv^\dagger(Y^\dagger Y)-\Lv^\dagger(Y^\dagger)Y-Y^\dagger\Lv^\dagger(Y)=\sum_Bw_B[B,Y]^\dagger[B,Y].
\label{eq:dissipation}
\end{equation}
The Hamiltonian part cancels on the left, and each dissipator leaves the
squared commutator of its jump with $Y$. Suppose $\Lv^\dagger Y=i\xi Y$ with
real $\xi$, so that $\Lv^\dagger Y^\dagger=-i\xi Y^\dagger$. Taking the
expectation of Eq.~\eqref{eq:dissipation} in a faithful stationary state makes
the left side vanish, so $[B,Y]=0$ for every jump. The dissipative part of
$\Lv^\dagger$ then annihilates $Y$, and $[H,Y]=\xi Y$. Now, for $x\in
V=\{x:[Y,c(x)]=0\}$,
$[Y,[H,c(x)]]=[[Y,H],c(x)]+[H,[Y,c(x)]]=-\xi[Y,c(x)]=0$, so $V$ is invariant
under $h^{\mathsf T}$, and likewise for creators. As in Step 1, $Y$ commutes
with every $c_a$ and $c_a^\dagger$, so $Y=zI$, and $[H,Y]=\xi Y$ forces
$\xi=0$ unless $Y=0$. Thus the kernel of $\Lv^\dagger$ is spanned by $I$, the
kernel of $\Lv$ is one-dimensional, and no nonzero imaginary eigenvalue exists.

\textit{Step 4: no growth.} The adjoint semigroup $e^{t\Lv^\dagger}$ is
unital and completely positive, so the Kadison--Schwarz inequality gives
$\|e^{t\Lv^\dagger}Y\|\le\|Y\|$ for all $t\ge0$. An eigenvalue with positive
real part, or a nontrivial Jordan block on the imaginary axis, would make
$e^{t\Lv^\dagger}$ grow. Hence zero is algebraically simple and every other
eigenvalue has negative real part, which gives Eq.~\eqref{eq:relax}.

The argument never restricts $Y$ to even operators, so it excludes persistent
coherences between Fock sectors of opposite parity as well. The decay rate it
implies depends on $n$, $h$, $\Gamma_{\rm L}$, $\Gamma_{\rm R}$ and $A$, and we
do not bound it uniformly.

\section{Exact current and variance at finite affinity}
\label{sec:cumulants}

The product formula of the main text separates partition noise from thermal
noise, and every bound in the paper is a bound on one of these two terms.
This section derives that formula from the full-Fock generator at finite
affinity. The counting field gives the current and variance as derivatives
of one eigenvalue. The two-point functions close, the stationary state is
Gaussian, and two Fourier identities reduce the variance to integrals of the
transmission.

\subsection{Counting statistics of a channel}

For any member of the class and counted channel $\mu$, the tilted generator is
$\Lv_s=\Lv+(e^s-1)\mathcal J_++(e^{-s}-1)\mathcal J_-$, with
$\mathcal J_+\rho=\gamma_\mu(n_\mu+1)L_\mu\rho L_\mu^\dagger$ and
$\mathcal J_-\rho=\gamma_\mu n_\mu L_\mu^\dagger\rho L_\mu$. Put
$\mathcal K_1=\mathcal J_+-\mathcal J_-$ and $\mathcal K_2=\mathcal J_++\mathcal
J_-$. Let zero be an algebraically simple eigenvalue of $\Lv$, with the rest of
the spectrum in the open left half-plane. Then the eigenvalue $\lambda(s)$ of
$\Lv_s$ continuing zero is analytic near $s=0$, and it is the scaled
cumulant generating function of the count. Expanding $\Lv_s\rho_s=\lambda(s)\rho_s$
with $\Tr\rho_s=1$ to second order in $s$ gives, with $\rho$ the stationary
state,
\begin{equation}
\begin{gathered}
j=\Tr(\mathcal K_1\rho),\qquad \Lv\eta=j\rho-\mathcal K_1\rho,\quad\Tr\eta=0,\\
d=\Tr(\mathcal K_2\rho)+2\Tr(\mathcal K_1\eta),
\end{gathered}
\label{eq:poisson}
\end{equation}
where $\eta=\partial_s\rho_s|_{s=0}$ is the first-order response of the
stationary state to the counting field. The trace of the first-order equation
gives $j=\lambda'(0)$, and the trace of the second-order equation gives
$d=\lambda''(0)$, the full variance rate with no factor $1/2$.

\subsection{Two-contact networks}

For the generator~\eqref{eq:gen} we count the right contact, extraction minus
injection, so $\mathcal J_+\rho=2\Gamma_{\rm R}f_{\rm L}c_n\rho c_n^\dagger$ and
$\mathcal J_-\rho=2\Gamma_{\rm R}f_{\rm R}c_n^\dagger\rho c_n$. Let
$G(E)=(E-h+i\Sigma)^{-1}$ and
\begin{equation}
T(E)=4\Gamma_{\rm L}\Gamma_{\rm R}|G_{n1}(E)|^2,\qquad
I_1=\int_{\mathbb R}T\frac{dE}{2\pi},\qquad
I_2=\int_{\mathbb R}T(1-T)\frac{dE}{2\pi},\qquad F_0=\frac{I_2}{I_1}.
\label{eq:Tnet}
\end{equation}
For the chain this is the transmission of the main text, with
$4\Gamma_{\rm L}\Gamma_{\rm R}=4\Gamma^2$.

The current and variance are then fixed by the two transmission integrals.
Under the hypotheses of Sec.~\ref{sec:unique}, and if $I_1>0$, the
transmission obeys $0\le T\le1$, $0<F_0<1$, and the right-contact count has
\begin{equation}
j=qI_1,\qquad d=q^2I_2+2f_{\rm L}f_{\rm R}I_1,\qquad \sigma=2Aj,
\label{eq:jd}
\end{equation}
so that
\begin{equation}
P=\frac{\sigma d}{j^2}=2AqF_0+\beta(A),\qquad \beta(A)=\frac{2A}{\sinh A}.
\label{eq:Pformula}
\end{equation}
The generator belongs to the coherently driven thermal-jump class with
$H_0=A(I-\hat n_1+\hat n_n)$ and $X=H-H_0$. Every chain with nonzero real
hoppings, real on-site energies and $\Gamma_{\rm L},\Gamma_{\rm R}>0$ satisfies
these hypotheses, with $I_1>0$ by Eq.~\eqref{eq:cofactor}.

For boundary-driven chains the formulas~\eqref{eq:jd} were obtained by Meier
et al.~\cite{Meier_2026eg} from the scattering description. The proof below
establishes that they give the long-time cumulants of the full-Fock
generator~\eqref{eq:gen}, for chains and for networks with complex hoppings
and unequal contacts. Equation~\eqref{eq:Pformula} shows the two noise
sources: partition noise, weighted by $2AqF_0$, and thermal noise,
$\beta(A)$, which decays only when the affinity grows.

\textit{Thermal class.} Take the forward channels $L_{\rm L}=c_1^\dagger$ and
$L_{\rm R}=c_n$, with $n_A=1/(e^A-1)$, $\gamma_\alpha=2\Gamma_\alpha q$ and
temperature one. Then $[H_0,c_1^\dagger]=-Ac_1^\dagger$ and
$[H_0,c_n]=-Ac_n$, so both gaps equal $A$, and
$\gamma_\alpha(n_A+1)=2\Gamma_\alpha f_{\rm L}$,
$\gamma_\alpha n_A=2\Gamma_\alpha f_{\rm R}$. These are the rates
of~\eqref{eq:gen}, and the class allows the drive $X$ to be chosen
independently of the baths. The total number $\sum_a\hat n_a$ commutes with
$H$; left forward jumps raise it and right forward jumps lower it, so the
stationary net currents of the two channels are equal. Each carries affinity
$A$, hence $\sigma=2Aj$.

\textit{Two-point closure.} For any operator $Y$ define its moment matrix
$C[Y]_{ab}=\Tr(c_b^\dagger c_aY)$. Inserting the adjoint dissipators,
$\Dis[c_p]^\dagger(c_b^\dagger c_a)=-\tfrac12(\delta_{pa}+\delta_{pb})c_b^\dagger c_a$
and
$\Dis[c_p^\dagger]^\dagger(c_b^\dagger c_a)=-\tfrac12(\delta_{pa}+\delta_{pb})c_b^\dagger c_a+\delta_{pa}\delta_{pb}I$,
together with $[H,c_b^\dagger c_a]$, gives for every operator $Y$
\begin{equation}
C[\Lv Y]=KC[Y]+C[Y]K^\dagger+Q_{\rm s}\Tr Y,\qquad
Q_{\rm s}=2\Gamma_{\rm L}f_{\rm L}\Pi_1+2\Gamma_{\rm R}f_{\rm R}\Pi_n .
\label{eq:closure}
\end{equation}
The Hamiltonian contributes $-i[h,C[Y]]$, each contact damps the row and
column of its mode at the total rate $2\Gamma_\alpha$, and the gain terms feed
$Q_{\rm s}$. The identity is linear in $Y$ and needs no positivity,
Gaussianity or parity of $Y$. With $U(t)=e^{Kt}$, strict stability makes $U$
decay exponentially, Jordan blocks included. Since
$\tfrac{d}{dt}(UU^\dagger)=-2U\Sigma U^\dagger$,
\begin{equation}
2\int_0^\infty U\Sigma U^\dagger\,dt=I .
\label{eq:unitarity-t}
\end{equation}
The stationary covariance $C\equiv C[\rho_*]$ solves
Eq.~\eqref{eq:closure} with left side zero. Writing
$Q_{\rm s}=2f_{\rm R}\Sigma+2q\Gamma_{\rm L}\Pi_1$ and using
Eq.~\eqref{eq:unitarity-t},
\begin{equation}
C=\int_0^\infty UQ_{\rm s}U^\dagger\,dt=f_{\rm R}I+q\Omega,\qquad
\Omega=2\Gamma_{\rm L}\int_0^\infty U\Pi_1U^\dagger\,dt .
\label{eq:covariance}
\end{equation}
Equation~\eqref{eq:unitarity-t} also gives $0\preceq\Omega\preceq I$, so the
eigenvalues of $C$ lie in $[f_{\rm R},f_{\rm L}]\subset(0,1)$: every mode is
partially filled at finite affinity, and $\Omega$ measures how much of the
left bath each mode sees.

\textit{Gaussian stationary state.} Quadratic Lindblad generators with linear
jumps have Gaussian stationary states~\cite{Prosen_2008tq}; for the present
generator we prove it directly. Diagonalize $C=O\operatorname{diag}(k_\ell)O^\dagger$
with $O$ unitary and $0<k_\ell<1$. With modes
$\tilde c_\ell=\sum_a\overline{O_{a\ell}}c_a$, define
\begin{equation}
\rho_G=\prod_\ell\big[k_\ell\,\tilde c_\ell^\dagger\tilde c_\ell+(1-k_\ell)\,\tilde c_\ell\tilde c_\ell^\dagger\big].
\label{eq:rhoG}
\end{equation}
It is positive definite with unit trace, its eigenvalues are products of
factors $k_\ell$ and $1-k_\ell$, and
$\Tr(c_b^\dagger c_a\rho_G)=\sum_\ell O_{a\ell}\overline{O_{b\ell}}k_\ell=C_{ab}$.
Because each normal mode is independent, its four-point functions obey Wick's
rule:
\begin{equation}
\Tr\big(c_b^\dagger c_a\,c_k^\dagger c_l\,\rho_G\big)-C_{ab}C_{lk}=C_{lb}(\delta_{ak}-C_{ak}).
\label{eq:fourpoint}
\end{equation}
The state $\rho_G$ is proportional to $\exp(\sum_{ab}c_a^\dagger J_{ab}c_b)$
with $J$ Hermitian. Conjugation by such an exponential maps each $c_a$ to a
linear combination of the $c_b$, and each $c_a^\dagger$ to one of the
$c_b^\dagger$. Hence every term of $\Lv\rho_G$ is a polynomial times
$\rho_G$: for instance
$c_n\rho_Gc_n^\dagger=c_n(\rho_Gc_n^\dagger\rho_G^{-1})\rho_G$. Each
polynomial is quadratic and conserves particle number, so after normal
ordering
\begin{equation}
\Lv\rho_G=\Big(\sum_{kl}Y_{kl}c_k^\dagger c_l+y\Big)\rho_G
\label{eq:Lrho}
\end{equation}
for some complex matrix $Y$ and number $y$. Trace preservation gives
$y=-\sum_{kl}Y_{kl}C_{lk}$, and Eq.~\eqref{eq:fourpoint} then gives
$C[\Lv\rho_G]=(I-C)YC$. Equation~\eqref{eq:closure} holds for every operator,
and its right side vanishes at $C[\rho_G]=C$, so $(I-C)YC=0$. Both $C$ and
$I-C$ are invertible, so $Y=0$, $y=0$, and $\Lv\rho_G=0$. By the uniqueness of
Sec.~\ref{sec:unique}, $\rho_G=\rho_*$. In particular $\rho_*$ is faithful because every
$k_\ell$ lies strictly between $0$ and $1$.

\textit{Current.} Let $v=e_n$, $x=Cv$ and $\omega=\Omega v$. From
Eq.~\eqref{eq:poisson},
$j=2\Gamma_{\rm R}[f_{\rm L}C_{nn}-f_{\rm R}(1-C_{nn})]=2\Gamma_{\rm R}(C_{nn}-f_{\rm R})=2\Gamma_{\rm R}q\Omega_{nn}$.
With the Fourier convention $\hat f(E)=\int e^{iEt}f(t)\,dt$, the stable
propagator obeys $\int_0^\infty e^{iEt}U(t)\,dt=iG(E)$. Let
$\varphi(t)=\langle n|U(t)|1\rangle$ for $t\ge0$ and $\varphi(t)=0$ for
$t<0$, so $|\hat\varphi|=|G_{n1}|$. Plancherel's theorem gives
$I_1=4\Gamma_{\rm L}\Gamma_{\rm R}\int_0^\infty|\varphi|^2dt=2\Gamma_{\rm
R}\Omega_{nn}$, hence $j=qI_1$.

\textit{Response.} Let $\Theta=C[\eta]$. Wick's rule gives the two contractions
\begin{equation}
C[c_n\rho_*c_n^\dagger]=C_{nn}C-xx^\dagger,\qquad
C[c_n^\dagger\rho_*c_n]=(1-C_{nn})C+(v-x)(v-x)^\dagger .
\label{eq:contractions}
\end{equation}
In the first, the direct pairing of $\langle c_n^\dagger c_b^\dagger c_ac_n\rangle$
gives $C_{nn}C_{ab}$ and the exchange pairing gives $-C_{an}C_{nb}$. Applying
Eq.~\eqref{eq:closure} to the response equation of~\eqref{eq:poisson}, with
$\Tr\eta=0$, the terms proportional to $C$ cancel because
$j=2\Gamma_{\rm R}(C_{nn}-f_{\rm R})$. Substituting $x=f_{\rm R}v+q\omega$, the
cross terms cancel as well, and
\begin{equation}
K\Theta+\Theta K^\dagger=2\Gamma_{\rm R}\big[f_{\rm L}xx^\dagger+f_{\rm R}(v-x)(v-x)^\dagger\big]
=2\Gamma_{\rm R}\big[q^2\omega\omega^\dagger+f_{\rm L}f_{\rm R}\Pi_n\big].
\label{eq:response}
\end{equation}
Its unique solution is
$\Theta=-2\Gamma_{\rm R}\int_0^\infty U[q^2\omega\omega^\dagger+f_{\rm L}f_{\rm R}\Pi_n]U^\dagger dt$.
The response is thus driven by two sources at the counted contact: the
bias-induced correlation $\omega$ and the thermal occupation noise
$f_{\rm L}f_{\rm R}$.

\textit{Variance.} Since $\Tr\eta=0$, $\Tr(\mathcal K_1\eta)=2\Gamma_{\rm
R}\Theta_{nn}$, and $\Tr(\mathcal K_2\rho_*)=2\Gamma_{\rm R}(f_{\rm
R}+qC_{nn})=4\Gamma_{\rm R}f_{\rm L}f_{\rm R}+q^2I_1$. Therefore
\begin{equation}
d=4\Gamma_{\rm R}f_{\rm L}f_{\rm R}+q^2I_1
-8\Gamma_{\rm R}^2\int_0^\infty\Big[q^2|\langle n|U(t)\omega\rangle|^2+f_{\rm L}f_{\rm R}|\langle n|U(t)|n\rangle|^2\Big]dt .
\label{eq:dintermediate}
\end{equation}
Two identities reduce the integrals. The $nn$ element of
Eq.~\eqref{eq:unitarity-t} reads
\begin{equation}
8\Gamma_{\rm R}^2\int_0^\infty|\langle n|U(t)|n\rangle|^2dt=4\Gamma_{\rm R}-2I_1 .
\label{eq:id1}
\end{equation}
For the other, let $\psi(t)=\int_{\mathbb R}\overline{\varphi(s)}\varphi(s+t)\,ds$.
For $t\ge0$ the semigroup law and Eq.~\eqref{eq:covariance} give
\begin{equation}
\langle n|U(t)\omega\rangle=2\Gamma_{\rm L}\int_0^\infty\langle n|U(t+s)|1\rangle\langle1|U(s)^\dagger|n\rangle\,ds=2\Gamma_{\rm L}\psi(t).
\label{eq:autocorr}
\end{equation}
This step keeps the ordered amplitude $\langle n|U|1\rangle$ throughout and
uses no reciprocity, so it holds for complex $h$. Since
$\psi(-t)=\overline{\psi(t)}$ and $\hat\psi=|\hat\varphi|^2=|G_{n1}|^2$,
Plancherel's theorem on the full line gives
\begin{equation}
\int_{\mathbb R}T^2\frac{dE}{2\pi}=16\Gamma_{\rm L}^2\Gamma_{\rm R}^2\int_{\mathbb R}|\psi|^2dt
=8\Gamma_{\rm R}^2\int_0^\infty|\langle n|U(t)\omega\rangle|^2dt .
\label{eq:id2}
\end{equation}
Inserting Eqs.~\eqref{eq:id1} and \eqref{eq:id2} into
Eq.~\eqref{eq:dintermediate}, the terms proportional to $\Gamma_{\rm R}$
cancel and $d=q^2(I_1-\int T^2dE/2\pi)+2f_{\rm L}f_{\rm R}I_1$, which is
Eq.~\eqref{eq:jd}. With $4f_{\rm L}f_{\rm R}/q=2/\sinh A$, the product
$P=2Ad/j$ becomes Eq.~\eqref{eq:Pformula}. Exponential decay of $U$ justifies
every exchange of integrals.

\textit{Range of $T$.} Let $W=(\sqrt{\Gamma_{\rm L}}e_1,\sqrt{\Gamma_{\rm
R}}e_n)$, so $WW^\dagger=\Sigma$, and
$\mathcal S(E)=I-2iW^\dagger G(E)W$. For real $E$ the resolvent identity
$G-G^\dagger=-2iG\Sigma G^\dagger$ makes $\mathcal S$ unitary, and
$|\mathcal S_{21}|^2=T$, so $0\le T\le1$. Strict stability removes real poles,
and $G_{n1}(E)=O(E^{-2})$ because the two contact modes are orthogonal, so
$T=O(E^{-4})$ is integrable. Since $1-T=|\mathcal S_{11}|^2$ and
$\mathcal S_{11}(E)\to1$ as $|E|\to\infty$, the rational function $1-T$ is not
identically zero, so $T<1$ except at finitely many energies; with $I_1>0$ this gives
$0<I_2<I_1$. For a chain, Eq.~\eqref{eq:cofactor} below shows $T>0$ on the
whole real axis, so $I_1>0$ holds automatically. This completes
Eqs.~\eqref{eq:jd} and \eqref{eq:Pformula}.

As a check of the normalization, the chain with $n=2$ and $t_1=\Gamma$ has
$I_1=\Gamma/2$ and $F_0=1/4$, so at $A=\log2$ the formulas give $j=\Gamma/6$,
$d=17\Gamma/72$ and $P=17\log2/6$. Under the common scaling
$(h,\Gamma_{\rm L},\Gamma_{\rm R})\mapsto\lambda(h,\Gamma_{\rm L},\Gamma_{\rm
R})$ with $\lambda>0$, $T_\lambda(E)=T(E/\lambda)$, so $I_1,I_2,j,d$ scale by
$\lambda$ while $F_0$ and $P$ are unchanged.

\section{Transmission of a chain and the lower bound on partition noise}
\label{sec:lower}

Partition noise lives where the transmission is neither near zero nor near
one, so its size is set by how steeply a finite chain can switch its
transmission. This section shows that the transmission of a chain is
$1/(1+Q)$ with $Q$ a polynomial of degree $2n$. Markov's inequality then
bounds how fast $Q$ can rise at every window edge. The result is the lower
bound on partition noise and on the product used in the main text.

\subsection{The transmission is $1/(1+Q)$}

In this section a chain may carry arbitrary real on-site energies
$\varepsilon_1,\dots,\varepsilon_n$, nonzero real hoppings
$t_1,\dots,t_{n-1}$ and contact linewidths $\Gamma_{\rm L},\Gamma_{\rm R}>0$
on modes $1$ and $n$; equal contacts are the case
$\Gamma_{\rm L}=\Gamma_{\rm R}=\Gamma$. A chain is a two-contact network whose
drift is strictly stable by Sec.~\ref{sec:stable}, so
Eqs.~\eqref{eq:jd} and \eqref{eq:Pformula} apply to it. Put
$h_-=h-i\Gamma_{\rm L}\Pi_1-i\Gamma_{\rm R}\Pi_n$,
$h_z=h+i\Gamma_{\rm L}\Pi_1-i\Gamma_{\rm R}\Pi_n$, $D(E)=\det(E-h_-)$,
$Z(E)=\det(E-h_z)$ and $\tau=\prod_{j=1}^{n-1}t_j$. Deleting the first row and
last column of $E-h_-$ leaves an upper triangular matrix: its entry in row
$l$ and column $l$ is $-t_l$, and every entry below the diagonal comes from a
pair of modes more than one site apart. The on-site energies and the
linewidths sit only on the diagonal of $E-h_-$, in positions that are either
deleted or above the diagonal of the minor. The cofactor sign cancels the
$n-1$ minus signs. Hence
\begin{equation}
G_{n1}(E)=\frac{\tau}{D(E)},
\label{eq:cofactor}
\end{equation}
with a numerator independent of energy, of the on-site energies and of the
linewidths. The left reflection amplitude
$\mathcal S_{11}=1-2i\Gamma_{\rm L}G_{11}$ follows from the matrix determinant
lemma applied to the rank-one change $E-h_z=E-h_--2i\Gamma_{\rm L}\Pi_1$:
\begin{equation}
\mathcal S_{11}(E)=\frac{Z(E)}{D(E)},\qquad |\mathcal S_{11}|^2=1-T .
\label{eq:reflection}
\end{equation}
Since $T=4\Gamma_{\rm L}\Gamma_{\rm R}|G_{n1}|^2$, unitarity gives
$|D(E)|^2=4\Gamma_{\rm L}\Gamma_{\rm R}\tau^2+|Z(E)|^2$ for real $E$.

The transmission is therefore a rational function with a constant numerator.
For every chain with nonzero hoppings, real on-site energies and
$\Gamma_{\rm L},\Gamma_{\rm R}>0$, and every real $E$,
\begin{equation}
T(E)=\frac1{1+Q(E)},\qquad Q(E)=\frac{Z(E)\overline{Z(\bar E)}}{4\Gamma_{\rm L}\Gamma_{\rm R}\tau^2},
\label{eq:TQ}
\end{equation}
where $Q$ is a real polynomial of degree exactly $2n$, with positive leading
coefficient $1/(4\Gamma_{\rm L}\Gamma_{\rm R}\tau^2)$ and $Q\ge0$ on the real
line. Indeed, by Eqs.~\eqref{eq:cofactor} and \eqref{eq:reflection},
$T=4\Gamma_{\rm L}\Gamma_{\rm R}\tau^2/(4\Gamma_{\rm L}\Gamma_{\rm R}\tau^2+|Z|^2)$. The product
$Z(E)\overline{Z(\bar E)}$ is invariant under conjugation of its coefficients,
equals $|Z(E)|^2$ for real $E$, and is monic of degree $2n$ because $Z$ is.

The polynomial $Q$ measures reflection relative to transmission. Its real zeros
are the energies of perfect transmission, and they may be degenerate. The
coefficients of $Z$ can be complex for a chain without mirror symmetry: for
$n=3$ with hoppings $(t_1,t_2)$, zero on-site energies and equal contacts,
$Z=E^3+(\Gamma^2-t_1^2-t_2^2)E+i\Gamma(t_2^2-t_1^2)$, so the conjugation in
Eq.~\eqref{eq:TQ} is needed. Equation~\eqref{eq:TQ} rests on the constant numerator of
Eq.~\eqref{eq:cofactor}; a general network has an energy-dependent cofactor
and need not have this form.

\subsection{Markov's inequality and the window edges}

Markov's inequality~\cite{Markov_1890oa} states that every real polynomial $p$
of degree at most $m$ obeys
\begin{equation}
\max_{[-1,1]}|p'|\le m^2\max_{[-1,1]}|p| ;
\label{eq:markov}
\end{equation}
see also the statement in Ref.~\cite{Aptekarev_2014ao}. A polynomial bounded
on an interval cannot rise faster than its degree squared allows.

Applied at the edges of every window where $Q\le1$, this bounds the
partition noise of every profile $1/(1+Q)$ from below. Let $Q$ be a real
polynomial, nonnegative on $\mathbb R$, of even degree $m\ge2$ with positive
leading coefficient, and $T=1/(1+Q)$. Then
\begin{equation}
\frac{\int_{\mathbb R}T(1-T)\,dx}{\int_{\mathbb R}T\,dx}\ge\frac2{9m^2}.
\label{eq:polyratio}
\end{equation}

Since $Q(x)\ge c|x|^m$ for large $|x|$ and some $c>0$, both
integrals converge, and the denominator is positive. Let $S=\{x:Q(x)\le1\}$.
The nonzero polynomial $Q-1$ has finitely many real zeros, so $S$ is a finite
union of disjoint closed intervals of positive length and finitely many
points; it is compact and possibly empty. Points have measure zero. Each
interval $I=[a,b]$ of length $\ell=b-a$ has $Q(a)=Q(b)=1$, since otherwise it
would extend, and $0\le Q\le1$ on $I$. Applying Eq.~\eqref{eq:markov} to
$p(y)=2Q(\tfrac{a+b}2+\tfrac\ell2y)-1$, which satisfies $|p|\le1$ on $[-1,1]$
and $p'(y)=\ell Q'(\tfrac{a+b}2+\tfrac\ell2y)$, gives
\begin{equation}
\max_I|Q'|\le\frac{m^2}{\ell}.
\label{eq:Qslope}
\end{equation}
Put $\delta=\ell/(2m^2)$. The strips $[a,a+\delta]$ and $[b-\delta,b]$ are
disjoint because $2\delta<\ell$, and by Eq.~\eqref{eq:Qslope} each has
$\tfrac12\le Q\le1$. There $T(1-T)=Q/(1+Q)^2\ge2/9$, since $Q/(1+Q)^2$
increases on $[0,1]$. Hence
\begin{equation}
\int_IT(1-T)\,dx\ge\frac29\cdot\frac{\ell}{m^2}\ge\frac2{9m^2}\int_IT\,dx,
\label{eq:edgestrips}
\end{equation}
using $T\le1$ in the last step. Each window edge thus carries partition noise
proportional to the window length divided by $m^2$. Off $S$, $Q>1$ and
$T(1-T)=TQ/(1+Q)\ge T/2\ge2T/(9m^2)$. Summing over the intervals and
integrating over the complement proves Eq.~\eqref{eq:polyratio}.

The argument divides by neither the length of $S$ nor the total
transmission, so it covers empty $S$, touching minima, degenerate zeros, and
several disconnected windows.

\subsection{No chain filters faster than $1/n^2$}
\label{sec:sizelower}

For a chain this gives the bound on partition noise of the main text, and
with the thermal term the bound on the product. Every chain with $n\ge2$
modes, nonzero hoppings, real on-site energies and
contact linewidths $\Gamma_{\rm L},\Gamma_{\rm R}>0$ obeys
\begin{equation}
F_0\ge\frac1{18n^2}.
\label{sm-eq:F0lower}
\end{equation}
For $n\ge3$ and every finite $A>0$,
\begin{equation}
P\ge\frac{\log n}{18n^2}.
\label{eq:Plower}
\end{equation}
No mirror symmetry, bound on coupling ratios, or current floor is assumed.

Equation~\eqref{eq:TQ} gives $T=1/(1+Q)$ with $m=2n$, and the
factor $1/2\pi$ cancels in $F_0$, so Eq.~\eqref{eq:polyratio} gives
Eq.~\eqref{sm-eq:F0lower}. For the product we split the affinity axis at
$A=\log n$ in Eq.~\eqref{eq:Pformula}. If $A\ge\log n$, then
$q\ge(n-1)/(n+1)\ge\tfrac12$ and $P\ge AF_0\ge\log n/(18n^2)$. If
$0<A\le\log n$, we use that $\beta$ is decreasing: its derivative has the sign
of $\sinh A-A\cosh A$, which vanishes at zero and has derivative
$-A\sinh A<0$. Hence
\begin{equation}
P\ge\beta(A)\ge\beta(\log n)=\frac{4n\log n}{n^2-1}\ge\frac{\log n}{18n^2},
\end{equation}
the last step being $72n^3\ge n^2-1$; this is Eq.~\eqref{eq:Plower}.

At small affinity the thermal term alone exceeds the bound, and at large
affinity the partition term alone does. Neither term can vanish without the
other growing, which is why the bound decays only as $\log n/n^2$.

\section{The Gegenbauer profile}
\label{sec:gegenbauer}

The lower bound asks for a profile whose small values spread over the window
and whose rise at the edges is nearly as steep as Markov's inequality
permits. This section shows that the normalized Gegenbauer polynomial has
both properties, so that its partition noise falls below $6/n^2$,
including both infinite tails.

For $n\ge1$ let
\begin{equation}
r_n(x)=\frac{2\Leg_{n+1}'(x)}{(n+1)(n+2)},\qquad T_n(x)=\frac1{1+r_n(x)^2},
\label{eq:rn}
\end{equation}
and $\mu_n=\int_{\mathbb R}T_n\,dx$, $\nu_n=\int_{\mathbb R}T_n(1-T_n)\,dx$.
The derivative formula for Jacobi polynomials~\cite[Eq.~18.9.15]{Olver_Book_NIST}
gives $r_n=\mathsf P^{(1,1)}_n/(n+1)$, and the relation to ultraspherical
polynomials~\cite[Eq.~18.7.1]{Olver_Book_NIST} gives
$r_n(x)=\mathsf C^{(3/2)}_n(x)/\mathsf C^{(3/2)}_n(1)$, with
$\mathsf C^{(\lambda)}_n$ the Gegenbauer polynomial. This normalized
Gegenbauer polynomial is a classical filter prototype~\cite{Stojanovic_2016dt};
the integral bounds below and their use for the uncertainty product are ours.

The transmitted weight of the profile stays close to the window length $2$,
and its partition noise is of order $n^{-2}$: for every integer $n\ge1$,
\begin{equation}
2-\frac4{(n+1)(n+2)}\le\mu_n\le2+\frac8{n(n+3)},\qquad
\frac{\nu_n}{\mu_n}\le B_G(n)=\frac6{n(n+3)}+\frac8{n^2(n+3)^2}<\frac6{n^2}.
\label{eq:BG}
\end{equation}
The derivation bounds the band and the tails separately.

\textit{Endpoint expansion.} The explicit Jacobi
series~\cite[Eq.~18.5.7]{Olver_Book_NIST} with $\alpha=\beta=1$ and $x=1+u$
gives, for every real $u$,
\begin{equation}
r_n(1+u)=\sum_{k=0}^n e_{nk}u^k,\qquad
e_{nk}=\frac{n!\,(n+k+2)!}{(n-k)!\,(n+2)!\,2^k\,k!\,(k+1)!}>0 .
\label{eq:endpoint}
\end{equation}
Thus $e_{n0}=1$, $e_{n1}=n(n+3)/4$, and the leading coefficient
$e_{nn}=(2n+2)!/[2^n(n+2)!(n+1)!]$ is positive, so $r_n^2$ has degree exactly
$2n$. Parity~\cite[Table~18.6.1]{Olver_Book_NIST} gives
$r_n(-x)=(-1)^nr_n(x)$, so $T_n$ is even. Every coefficient being positive
means the profile rises at least linearly outside the window, at rate
$n(n+3)/4$, all the way to infinity.

\textit{In-band norm.} Put $m=n+1$. From $\Leg_m(\pm1)=(\pm1)^m$ and
$\Leg_m'(\pm1)=(\pm1)^{m+1}m(m+1)/2$, and since $\Leg_m''$ has degree below $m$
and is orthogonal to $\Leg_m$ with unit weight, integration by parts gives
$\int_{-1}^1(\Leg_m')^2dx=[\Leg_m\Leg_m']_{-1}^1=m(m+1)$. Hence
\begin{equation}
\int_{-1}^1r_n^2\,dx=\frac4{(n+1)(n+2)}\equiv a_n .
\label{eq:inband}
\end{equation}
The ripple inside the window has mean square of order $n^{-2}$. With
$(1+Q)^{-1}\ge1-Q$ and $Q/(1+Q)^2\le Q$ pointwise for $Q=r_n^2$,
$2-a_n\le\int_{-1}^1T_n\le2$ and $\int_{-1}^1T_n(1-T_n)\le a_n$.

\textit{Both tails.} With $b_n=n(n+3)/4$, Eq.~\eqref{eq:endpoint} gives
$r_n(1+u)\ge1+b_nu$ for all $u\ge0$, so
\begin{equation}
\int_1^\infty T_n\,dx\le\int_0^\infty\frac{du}{(1+b_nu)^2}=\frac1{b_n}.
\label{eq:tail}
\end{equation}
The whole out-of-band mass is thus of order $n^{-2}$, with no cutoff. By
evenness and $T_n(1-T_n)\le T_n$, both tails together contribute at most
$2/b_n=8/[n(n+3)]$ to $\mu_n$ and to $\nu_n$.

\textit{Band and tails together.} Adding band and tails gives the bounds
on $\mu_n$ and $\nu_n\le a_n+8/[n(n+3)]$. With $z=n(n+3)$ and
$(n+1)(n+2)=z+2$,
$\nu_n/\mu_n\le[4/(z+2)+8/z]/[2-4/(z+2)]=(6z+8)/z^2=B_G(n)$, and
$6/n^2-B_G(n)=(18n+46)/[n^2(n+3)^2]>0$, which is Eq.~\eqref{eq:BG}.

For a common scale $w>0$, the transmission $T_n(E/w)$ has
$\int T_n(E/w)\,dE=w\mu_n$, and $F_0=\nu_n/\mu_n$ is unchanged. At $n=1$,
$T_1=(1+x^2)^{-1}$ gives $\mu_1=\pi$ and $\nu_1/\mu_1=1/2<B_G(1)=2$; the next
profiles are $r_2=(5x^2-1)/4$ and $r_3=(7x^3-3x)/4$.

\textit{Where the ripple and the edge are.} On the window $|r_n|\le1$, with
equality at $x=\pm1$~\cite[\S18.14]{Olver_Book_NIST}, so the ripple reaches its
largest value at the edges. Its mean square~\eqref{eq:inband} is nevertheless of
order $n^{-2}$, so the bulk transmits almost perfectly. At the edge the slope
of $Q=r_n^2$ is $Q'(1)=n(n+3)/2$, within a factor $4$ of the Markov
bound~\eqref{eq:Qslope}, $d^2/\ell=2n^2$ for $d=2n$ and $\ell=2$. The profile
therefore leaves the window within a factor $4$ of the steepest slope a
polynomial of its degree can have, and its noise has the order $n^{-2}$ of
Eq.~\eqref{eq:polyratio}. The
logarithm-free order is sharp for the scalar inequality: $Q=r_n^2$ has degree
$d=2n$ and $d^2\nu_n/\mu_n\le4n^2B_G(n)<24$, so no bound of the form
$d^2F_0\ge c\log d$ holds for all nonnegative polynomials.

\section{Exact realization by an equal-contact chain at every length}
\label{sec:realization}

The profile~\eqref{eq:rn} becomes physical only if a chain with equal contacts
transmits exactly $T_n(E/w)$. The construction below inverts the two-contact
scattering problem: the target fixes the denominator $D$ of
Eq.~\eqref{eq:cofactor} through a spectral factorization, the denominator fixes
an $n$-point spectral measure, and the measure fixes the Jacobi matrix.

\textit{Algorithm.} Fix $n\ge2$ and work at $w=1$.
\begin{enumerate}
\item Let $\kappa=1/e_{nn}$ and $M(z)=\kappa r_n(z)$, the monic polynomial
proportional to the target; $M$ has real coefficients and parity $n$.
\item The polynomial $F=M^2+\kappa^2$ is positive on the real line and has real
coefficients, so it has exactly $n$ roots, with multiplicity, in each open
half-plane. Let $D$ be the monic polynomial whose roots are those in the lower
half-plane, and write $D=D_1+iD_2$ with $D_1,D_2$ of real coefficients.
\item Put $\Gamma=\tfrac12[z^{n-1}]D_2$, $p=(D_1+M)/2$ and $q_D=D_2/(2\Gamma)$.
\item Let $x_1<\dots<x_n$ be the roots of $p$ and
$\rho_s=\kappa/(2\Gamma|p'(x_s)|)$.
\item Let $h$ be the Jacobi matrix of the measure $\sum_s\rho_s\delta_{x_s}$:
Gram--Schmidt orthonormalize $1,x,\dots,x^{n-1}$ in this measure with positive
leading coefficients, and take the matrix of multiplication by $x$.
\item Multiply $h$ and $\Gamma$ by $w$.
\end{enumerate}

The algorithm always succeeds, and the chain it returns transmits the
target exactly. For every integer $n\ge2$ and every $w>0$, the algorithm
terminates and
outputs a real $n\times n$ Jacobi matrix with zero diagonal and positive
hoppings $t_j=t_{n-j}$, and a contact $\Gamma>0$, such that the equal-contact
chain transmits
\begin{equation}
4\Gamma^2\big|\big(E-h+i\Gamma(\Pi_1+\Pi_n)\big)^{-1}_{n1}\big|^2=\frac1{1+r_n(E/w)^2}\qquad\text{for all real }E.
\label{eq:Tn-exact}
\end{equation}
Moreover $\det(E-h+i\Gamma(\Pi_1+\Pi_n))=D(E)$ and
$\prod_jt_j=\kappa/(2\Gamma)$ at $w=1$.

Steps (a) to (g) below derive this. By the scaling remark of
Sec.~\ref{sec:cumulants}, step 6 maps $T(E)$ to $T(E/w)$, so it suffices to
take $w=1$.

\textit{(a) Spectral factor.} The factorization $DD^\#=F$, with
$D^\#(z)=\overline{D(\bar z)}=D_1-iD_2$, reads
\begin{equation}
D_1^2+D_2^2=M^2+\kappa^2 .
\label{eq:spectralfactor}
\end{equation}
The polynomial $D_1$ is monic of degree $n$ and $D_2$ has degree at most
$n-1$. If the lower roots are $u_j-iv_j$ with $v_j>0$, then
$[z^{n-1}]D_2=\sum_jv_j>0$, so $\Gamma>0$ and $q_D$ is monic of degree $n-1$.

\textit{(b) Two Blaschke channels.} Define
\begin{equation}
\mathcal S(z)=\frac1{D(z)}\begin{pmatrix}M(z)&-i\kappa\\-i\kappa&M(z)\end{pmatrix},
\label{eq:Starget}
\end{equation}
the scattering matrix the chain must have. Its eigenvalues on the constant
vectors $(1,\mp1)/\sqrt2$ are $s_\pm=N_\pm/D$ with $N_\pm=M\pm i\kappa$. The
monic polynomials $N_\pm$ have no real roots, no common roots, and
$N_-=N_+^\#$. Let $L_+$ and $U_+$ be the lower and upper roots of $N_+$; the
lower roots of $N_-$ are then the conjugates of $U_+$, so
$D=\prod_{L_+}(z-\lambda)\prod_{U_+}(z-\bar\alpha)$ and
\begin{equation}
s_+(z)=\prod_{\alpha\in U_+}\frac{z-\alpha}{z-\bar\alpha},
\label{eq:blaschke}
\end{equation}
and likewise $s_-$ with $U_-$ the upper roots of $N_-$. The two sets have
$|U_+|+|U_-|=n$. Each channel is thus a finite Blaschke product, of modulus
one on the real axis. On the real axis write $s=e^{i\theta}$ with
$\theta(-\infty)=0$; differentiating each factor gives
\begin{equation}
\theta'(x)=\sum_{\alpha\in U}\frac{2\operatorname{Im}\alpha}{(x-\operatorname{Re}\alpha)^2+(\operatorname{Im}\alpha)^2}>0,\qquad
\theta(+\infty)=2\pi|U| .
\label{eq:phase}
\end{equation}
The phase winds monotonically, so $s=-1$ at exactly $|U|$ distinct real
points, where $\theta=(2j-1)\pi$. There are no other solutions: after
clearing denominators, $1+s$ has a numerator of degree $|U|$ with leading
coefficient $2$. The Cayley function $k=(1/i)(1-s)/(1+s)=-\tan(\theta/2)$ has
a simple pole at each such point $x_0$; near it,
$\theta=\pi+\theta'(x_0)(x-x_0)+\dots$, so
\begin{equation}
\operatorname{Res}_{z=x_0}k=\frac{2}{\theta'(x_0)}>0 .
\label{eq:residue}
\end{equation}
The two channels never share a pole, because $s_+-s_-=2i\kappa/D\ne0$ on the
real axis.

\textit{(c) The Cayley matrix.} The matrix
$\mathsf K=(1/i)(I-\mathcal S)(I+\mathcal S)^{-1}$ has eigenvalues $k_\pm$ on
the same vectors, hence $n$ distinct simple real poles, and its diagonal
residues are half the scalar residues~\eqref{eq:residue}. By
Eq.~\eqref{eq:spectralfactor} and direct $2\times2$ multiplication,
\begin{equation}
\mathsf K(z)=\frac1{D_1(z)+M(z)}\begin{pmatrix}D_2(z)&\kappa\\\kappa&D_2(z)\end{pmatrix}
=\frac1{2p(z)}\begin{pmatrix}2\Gamma q_D(z)&\kappa\\\kappa&2\Gamma q_D(z)\end{pmatrix}.
\label{eq:Kmatrix}
\end{equation}
For instance, the eigenvalue on $(1,1)/\sqrt2$, $(D_2+\kappa)/(D_1+M)$, equals
$k_-$ precisely because $(D_1-M)(D_1+M)=\kappa^2-D_2^2$. The off-diagonal
entry $\kappa/(2p)$ has a pole at every root of $p$, and $p$ is monic of
degree $n$, so $p$ has exactly $n$ simple real roots $x_1<\dots<x_n$, the poles
of $\mathsf K$.

\textit{(d) A positive spectral measure.} The entry $\mathsf
K_{11}=\Gamma q_D/p$ has positive residues by step (c), so
$\rho_s=q_D(x_s)/p'(x_s)>0$. Since $q_D/p=1/z+O(z^{-2})$, partial fractions
give $\sum_s\rho_s=1$. At a root of $p$, $D_1=-M$, and
Eq.~\eqref{eq:spectralfactor} gives $D_2(x_s)^2=\kappa^2$, so
\begin{equation}
\rho_s=\frac{\kappa}{2\Gamma|p'(x_s)|}>0,\qquad\sum_s\rho_s=1 .
\label{eq:measure}
\end{equation}
The target has thus produced a probability measure on $n$ distinct real
points.

\textit{(e) The Jacobi matrix.} A probability measure on $n$ distinct real
points with positive weights determines a unique real Jacobi matrix $h$ with
positive off-diagonal entries such that $\det(z-h)=\prod_s(z-x_s)$ and
$(z-h)^{-1}_{11}=\sum_s\rho_s/(z-x_s)$~\cite{Arlinskii_2006ns}. Step 5 is that
construction. No polynomial of degree below $n$ vanishes at all $n$ support
points, so Gram--Schmidt never breaks down. The isometry from $L^2(\rho)$
onto $\mathbb C^n$ that sends the orthonormal polynomials to the basis maps
multiplication by $x$ to $h$ and the constant $1$ to $e_1$. Here
$\det(z-h)=p(z)$ and $G^0_{11}=q_D/p$, where $G^0(z)=(z-h)^{-1}$. The parity
of $p$ is $n$: the lower roots of the even polynomial $F$ are invariant under
$\lambda\mapsto-\bar\lambda$, so $D(-z)=(-1)^nD^\#(z)$, and
$D_1(-z)=(-1)^nD_1(z)$. Hence the roots of $p$ are symmetric, opposite roots
carry equal weights by Eq.~\eqref{eq:measure}, the orthonormal polynomials
alternate in parity, and every diagonal entry $\int x\phi_j(x)^2d\rho$
vanishes.

\textit{(f) The full boundary resolvent.} Let $v_s$ be the real orthonormal
eigenvectors of $h$. Then $\rho_s=v_s(1)^2$, and the residues of
$G^0_{n1}=\tau/p$, from Eq.~\eqref{eq:cofactor} at $\Gamma=0$, give
$v_s(1)v_s(n)=\tau/p'(x_s)$. Hence, by Eq.~\eqref{eq:measure},
\begin{equation}
v_s(n)^2=\frac{\tau^2}{p'(x_s)^2\rho_s}=\Big(\frac{2\Gamma\tau}{\kappa}\Big)^2\rho_s .
\end{equation}
Summing over $s$, $1=(2\Gamma\tau/\kappa)^2$, so $\tau=\kappa/(2\Gamma)$ and
$v_s(n)^2=\rho_s$. The spectral measures of $h$ at $e_n$ and $e_1$ coincide,
so $G^0_{nn}=G^0_{11}$. The reversed matrix $JhJ$, with $J$ the reversal of
the basis, is a Jacobi matrix with positive off-diagonal entries whose measure
at $e_1$ is $\rho$; by uniqueness $JhJ=h$, that is, $t_j=t_{n-j}$. With
Eq.~\eqref{eq:Kmatrix}, the undamped boundary block is
\begin{equation}
G^0_b(z)=\begin{pmatrix}G^0_{11}&G^0_{1n}\\G^0_{n1}&G^0_{nn}\end{pmatrix}
=\frac1{2\Gamma p(z)}\begin{pmatrix}D_2(z)&\kappa\\\kappa&D_2(z)\end{pmatrix}=\frac{\mathsf K(z)}{\Gamma}.
\label{eq:Gb}
\end{equation}

\textit{(g) Equal contacts.} Let $W_1=(e_1,e_n)$ and
$\tilde G=(z-h+i\Gamma W_1W_1^\dagger)^{-1}$. The Woodbury identity gives
$\tilde G_b=G^0_b(I+i\Gamma G^0_b)^{-1}$, hence
\begin{equation}
I-2i\Gamma\tilde G_b=(I-i\mathsf K)(I+i\mathsf K)^{-1}=\mathcal S .
\end{equation}
The damped chain therefore has exactly the scattering matrix~\eqref{eq:Starget}.
Its off-diagonal entry gives $\tilde G_{n1}=\kappa/(2\Gamma D)=\tau/D$. The
damped cofactor is still $\tau$, so Eq.~\eqref{eq:cofactor} gives
$\det(z-h+i\Gamma(\Pi_1+\Pi_n))=D(z)$, and for real $E$
\begin{equation}
4\Gamma^2|\tilde G_{n1}(E)|^2=\frac{\kappa^2}{|D(E)|^2}=\frac{\kappa^2}{M(E)^2+\kappa^2}=\frac1{1+r_n(E)^2}.
\end{equation}
This proves Eq.~\eqref{eq:Tn-exact}.

The two contacts sit on distinct modes because $n\ge2$; the contact
linewidth $\Gamma$ is fixed by the target as half the subleading coefficient
of $D_2$, not chosen. For $n=2$, $r_2(E)=(5E^2-1)/4$, $M=E^2-1/5$ and
$\kappa=4/5$. The two-site chain has $D=(E+i\Gamma)^2-t_1^2$, and
$|D|^2=M^2+\kappa^2$ requires $t_1^2-\Gamma^2=1/5$ and
$t_1^2+\Gamma^2=\sqrt{17}/5$. Restoring $w$,
\begin{equation}
t_1^2=\frac{w^2(\sqrt{17}+1)}{10},\qquad\Gamma^2=\frac{w^2(\sqrt{17}-1)}{10},
\end{equation}
and indeed $2\Gamma t_1=\kappa w^2$. The shortest even chain is included; the
construction is not confined to odd lengths.

\section{The size law}
\label{sec:sizelaw}

The lower bound and the exact Gegenbauer chain together fix the minimal
product of an $n$-mode chain up to a constant factor, at a current that
stays finite. Let $\Pmin(n)$ be the infimum of $P$ over all chains with $n$ modes, nonzero
real hoppings, real on-site energies and contact linewidths
$\Gamma_{\rm L},\Gamma_{\rm R}>0$, and over all finite $A>0$. The lower bound
of Sec.~\ref{sec:sizelower} covers this class, and the chain of
Sec.~\ref{sec:realization}, with zero on-site energies and equal
contacts, lies inside it.

For every $n\ge2$ and $w>0$, the chain of Sec.~\ref{sec:realization}
at affinity $A_n=2\log n$ has
\begin{equation}
P_n\le V_n\equiv4q_n(\log n)B_G(n)+\frac{8\log n}{n^2(1-n^{-4})}<33\frac{\log n}{n^2},\qquad
q_n=\frac{n^2-1}{n^2+1},
\label{eq:Vn}
\end{equation}
and its current obeys
\begin{equation}
\frac w\pi q_n\Big(1-\frac2{(n+1)(n+2)}\Big)\le j_n\le\frac w\pi q_n\Big(1+\frac4{n(n+3)}\Big),\qquad
j_n\ge\frac{w}{2\pi},\qquad\Big|\frac{\pi j_n}{w}-1\Big|\le\frac4{n^2}.
\label{eq:current}
\end{equation}
Consequently, for every $n\ge3$,
\begin{equation}
\frac1{18n^2}\le\inf F_0<\frac6{n^2},\qquad
\frac{\log n}{18n^2}\le\Pmin(n)<33\frac{\log n}{n^2},
\label{sm-eq:sizelaw}
\end{equation}
the first pair holding also at $n=2$, with the infimum over the same chains.

The chain belongs to the class, so Eq.~\eqref{eq:Pformula}
applies. At $A_n=2\log n$, $q_n=\tanh(\log n)=(n^2-1)/(n^2+1)$ and
$\sinh(2\log n)=(n^2-n^{-2})/2$, so Eq.~\eqref{eq:Pformula} reads
$P_n=4q_n(\log n)F_0+8\log n/[n^2(1-n^{-4})]$, the thermal term being exact.
Equation~\eqref{eq:BG} gives $F_0\le B_G(n)<6/n^2$, hence
$P_n\le V_n$. With $0<q_n<1$ and $1-n^{-4}\ge15/16$,
$V_n<(24+128/15)\log n/n^2=(488/15)\log n/n^2<33\log n/n^2$. For the current,
$I_1=(w/2\pi)\mu_n$ and $j_n=q_nI_1$, so the bounds on $\mu_n$ give the first
pair in Eq.~\eqref{eq:current}. Both factors of the lower bound increase with
$n$, and at $n=2$ their product is $(3/5)(5/6)=1/2$. The deficit of the lower
bound is $2/(n^2+1)+2q_n/[(n+1)(n+2)]\le4/n^2$, and the excess of the upper
bound is at most $4/[n(n+3)]\le4/n^2$. Equation~\eqref{sm-eq:sizelaw} combines
these with Eqs.~\eqref{sm-eq:F0lower} and \eqref{eq:Plower}.

The chain reaches the order $\log n/n^2$ of the lower bound while its current
stays near $w/\pi$. The upper bound itself behaves as $V_n\sim32\log n/n^2$,
which is not a statement about the asymptotic coefficient of the true
product, and neither constant is claimed optimal. Every coefficient of the
chain is finite, positive and computable exactly for each $n$, but we prove
no bound on its coupling ratios $\max(\Gamma,t_j)/\min(\Gamma,t_j)$ as $n$
grows. The common scaling preserves $F_0$, $P$ and every coupling ratio, so it
cannot supply such a bound.

\section{The contact budget of two-contact networks}
\label{sec:budget}

Near-perfect transmission over a window spends a resource that the contacts
supply in fixed amount. This section shows that the weaker contact linewidth
caps the integrated logarithm of the reflection probability. It then
converts that cap into a bound on partition noise at given current. Let $h$
be Hermitian, $\Gamma_{\rm L},\Gamma_{\rm R}>0$, $K$ strictly stable
and $I_1>0$, and put $\Gamma_*=\min(\Gamma_{\rm L},\Gamma_{\rm R})$. Let
$z_1,\dots,z_n$ be the eigenvalues, with algebraic multiplicity, of
$h_z=h+i\Gamma_{\rm L}\Pi_1-i\Gamma_{\rm R}\Pi_n$. Then
\begin{equation}
\mathcal B\equiv\int_{\mathbb R}-\log\big(1-T(E)\big)\frac{dE}{2\pi}=\Gamma_{\rm L}+\Gamma_{\rm R}-\sum_{k=1}^n|\operatorname{Im}z_k|\le2\Gamma_*,
\label{eq:poleszeros}
\end{equation}
and
\begin{equation}
I_1\,\frac{\log(1/F_0)}{1-F_0}\le\mathcal B\le2\Gamma_* .
\label{sm-eq:budget}
\end{equation}

The logarithm of the reflection probability is the budget that near-perfect
transmission spends. It is fixed by the positions of the poles and zeros of
the reflection amplitude, the logarithmic method of Bode and
Fano~\cite{Fano_1950tl}, and it is capped by the weaker contact. The left
inequality converts the budget into a bound on partition noise at given
transmitted current.

\textit{Reflection determinant.} Let $h_-=h-i\Sigma$, whose eigenvalues
$p_k$ have $\operatorname{Im}p_k<0$ by strict stability. The matrix
determinant lemma applied to $E-h_z=E-h_--2i\Gamma_{\rm L}\Pi_1$ gives the
left reflection amplitude
\begin{equation}
\mathcal S_{11}(E)=1-2i\Gamma_{\rm L}G_{11}(E)=\frac{\det(E-h_z)}{\det(E-h_-)}=\frac{\prod_k(E-z_k)}{\prod_k(E-p_k)},
\label{eq:refl-network}
\end{equation}
and unitarity of $\mathcal S$ (Sec.~\ref{sec:cumulants}) gives
$|\mathcal S_{11}|^2=1-|\mathcal S_{21}|^2=1-T$. Both determinants are monic
of degree $n$. Common factors, real zeros and repeated roots are kept with
their multiplicities; no minimal realization is needed.

\textit{Convergence.} At large $|E|$, $T=O(E^{-4})$, so
$-\log(1-T)=O(E^{-4})$. The numerator of~\eqref{eq:refl-network} is a monic
polynomial, so its real zeros are finitely many, and near a
real zero of order $m$, $-\log|\mathcal S_{11}|^2=-2m\log|E-a|+O(1)$ is
integrable. Thus $\mathcal B$ is an ordinary Lebesgue integral with a
nonnegative integrand.

\textit{Pole--zero identity.} For real $a,b$, the antiderivative of
$\log[(E-a)^2+b^2]$ gives
\begin{equation}
\int_{-L}^L\log|E-a-ib|^2dE=4L\log L-4L+2\pi|b|+o(1)\qquad(L\to\infty),
\label{eq:rootint}
\end{equation}
including $b=0$. Applied to the $n$ zeros and $n$ poles
of~\eqref{eq:refl-network}, the divergent terms cancel, and monotone
convergence identifies the symmetric limit with $\mathcal B$:
$\mathcal B=\sum_k|\operatorname{Im}p_k|-\sum_k|\operatorname{Im}z_k|$. All
$p_k$ lie in the lower half-plane and $\operatorname{Im}\Tr h_-=-(\Gamma_{\rm
L}+\Gamma_{\rm R})$, so the first sum is $\Gamma_{\rm L}+\Gamma_{\rm R}$ for
any finite matrix. For the zeros,
\begin{equation}
\sum_k|\operatorname{Im}z_k|\ge\Big|\operatorname{Im}\sum_kz_k\Big|=|\operatorname{Im}\Tr h_z|=|\Gamma_{\rm L}-\Gamma_{\rm R}|,
\end{equation}
so $\mathcal B\le\Gamma_{\rm L}+\Gamma_{\rm R}-|\Gamma_{\rm L}-\Gamma_{\rm
R}|=2\Gamma_*$. This step replaces the total linewidth by the weaker one: a
strong contact cannot pay for a weak one.

\textit{Jensen step.} The measure $d\mu=T\,dE/(2\pi I_1)$ is a probability
measure, and $x=1-T$ has mean $\int x\,d\mu=F_0$. The function
\begin{equation}
g(x)=\frac{\log(1/x)}{1-x}=\int_0^1\frac{ds}{x+(1-x)s},\qquad0<x\le1,
\label{eq:gfun}
\end{equation}
is decreasing and convex, as its integral representation shows, with
$g(1)=1$. Its singular point $x=0$ occurs only at the finitely many energies of
perfect transmission, a $\mu$-null set. Jensen's inequality gives
$g(F_0)\le\int g(1-T)\,d\mu=\mathcal B/I_1$, which is
Eq.~\eqref{sm-eq:budget}.

Among measurable transmissions with given $I_1$ and $\mathcal B$,
Eq.~\eqref{sm-eq:budget} is the best possible relation: the flat transmission
$T=1-f$ on an interval of length $2\pi I_1/(1-f)$ and zero elsewhere has
$F_0=f$ and $\mathcal B=I_1g(f)$. Such a transmission is a relaxation, not a
network: a finite chain has $T>0$ everywhere. For the chain the bound reads
$\mathcal B=2\Gamma-\sum_k|\operatorname{Im}z_k|$. For the two-site chain with
$t_1<\Gamma$, $Z=E^2+\Gamma^2-t_1^2$ has zeros at $\pm
i\sqrt{\Gamma^2-t_1^2}$, so $\mathcal B=2\Gamma-2\sqrt{\Gamma^2-t_1^2}<2\Gamma$;
the budget is saturated only when the reflection zeros are real.

\section{The minimal contact cost}
\label{sec:contactcost}

The budget prices partition noise in contact linewidth, and the thermal term
fixes how much partition noise a target product allows. This section
combines the two into the minimal contact linewidth per transferred
particle, and shows that an equal-contact Chebyshev chain attains it up to an
additive constant.

\subsection{The scalar bound}

For $0<p<2$ let $a_p>0$ solve $\beta(a_p)=p$; it exists uniquely because
$\beta$ decreases strictly from $2$ to $0$ (Sec.~\ref{sec:lower}). For
$A>a_p$ define
\begin{equation}
f_p(A)=\frac{p-\beta(A)}{2Aq(A)},\qquad
H_p(A)=\frac{g(f_p(A))}{2q(A)},\qquad
\mathcal R(p)=\inf_{A>a_p}H_p(A),
\label{eq:Rp}
\end{equation}
with $q(A)=\tanh(A/2)$ and $g$ from Eq.~\eqref{eq:gfun}. The partition-noise
allowance left by the thermal term at affinity $A$ is $f_p(A)$, and $H_p(A)$
is the contact cost that the budget charges for it.

No network can pay less than $\mathcal R(p)$. Every network satisfying the
hypotheses of Eq.~\eqref{eq:poleszeros}, at any finite $A>0$, with
$P\le p<2$, obeys
\begin{equation}
\frac{\Gamma_*}{j}\ge\mathcal R(p).
\label{eq:costlower}
\end{equation}
The infimum in Eq.~\eqref{eq:Rp} is attained at a finite $A$, and for
$s=\log(1/p)\ge8$ it is squeezed as
\begin{equation}
0\le2\mathcal R(p)-\big[s+\log(2s)\big]\le\frac{2\log s+\log8+5}{s}.
\label{eq:squeeze}
\end{equation}

\textit{The bound and its attainment.} By Eq.~\eqref{eq:Pformula},
$P\le p$ requires $\beta(A)<p$, so $A>a_p$, and then $0<F_0\le f_p(A)$.
Also $f_p(A)<1$, because $\beta(A)+2Aq(A)=2A\coth A>2>p$. The function $g$
decreases, $g'(f)=[1-1/f-\log f]/(1-f)^2<0$, so Eq.~\eqref{sm-eq:budget} and
$j=qI_1$ give
\begin{equation}
\frac{\Gamma_*}{j}\ge\frac{g(F_0)}{2q}\ge\frac{g(f_p(A))}{2q}=H_p(A)\ge\mathcal R(p).
\end{equation}
As $A\downarrow a_p$, $f_p\to0$ and $H_p\to\infty$; as $A\to\infty$,
$f_p\sim p/(2A)$ and $H_p\to\infty$ again. By continuity the infimum is a
minimum.

\textit{Lower half of Eq.~\eqref{eq:squeeze}.} Let $s\ge8$ and $A>a_p$. Since
$\beta(1)>1>p$ and $\beta$ decreases, $A>1$; then
$\beta(A)=4Ae^{-A}/(1-e^{-2A})>e^{-A}$, so $A>s$. Hence $q>1/2$ and
$f=f_p(A)<p/A<e^{-1}$. With $Y=\log(1/f)>1$ and $\log(1/q)\le1/q-1$,
\begin{equation}
\frac{g(f)}q\ge\frac Yq\ge Y+\log\frac1q=\log\frac1{fq}\ge s+\log(2A)>s+\log(2s),
\end{equation}
since $fq\le p/(2A)$. Minimizing over $A$ gives the lower half. Small
precision forces an affinity at least $s$, and dividing the partition-noise
allowance by that affinity is what produces the $\log s$ correction.

\textit{Upper half of Eq.~\eqref{eq:squeeze}.} Take
$A_t=s+2\log s+\log8$, so $e^{-A_t}=p/(8s^2)$ and $s<A_t\le2s$ for $s\ge8$.
Then $\alpha\equiv\beta(A_t)/p=A_t/[2s^2(1-e^{-2A_t})]\le2/s\le1/4$, and
$f_t=f_p(A_t)=p(1-\alpha)/(2A_tq_t)$. With $X=s+\log(2s)$,
\begin{equation}
\log\frac1{f_t}=s+\log(2A_t)+\log q_t-\log(1-\alpha)\le X+\frac{2\log s+\log8+4}{s}\equiv X+\Delta,
\end{equation}
using $\log(A_t/s)\le(2\log s+\log8)/s$, $\log q_t\le0$ and
$-\log(1-\alpha)\le\alpha/(1-\alpha)\le4/s$. Next $q_t\ge1-2e^{-A_t}$ and
$f_t\le p/s$ give $q_t(1-f_t)\ge1-2p/s$, so $1/[q_t(1-f_t)]\le1+4p/s$.
Since $X+\Delta\le2s$ and $8p\le1/s$ for $s\ge8$,
\begin{equation}
2\mathcal R(p)\le2H_p(A_t)\le(X+\Delta)\Big(1+\frac{4p}s\Big)\le X+\Delta+8p\le X+\frac{2\log s+\log8+5}{s}.
\end{equation}

Equation~\eqref{eq:squeeze} gives
$\mathcal R(p)=\tfrac12\log(1/p)+\tfrac12\log\log(1/p)+\tfrac12\log2+O(\log s/s)$.
It is the exact optimum of the scalar relaxation: assigning the flat
transmission of Sec.~\ref{sec:budget}, with $f=f_p(A)$ and an interval of
normalized length $2\Gamma_*/\log(1/f)$, to the formulas~\eqref{eq:jd}
gives $P=p$ and $\Gamma_*/j=H_p(A)$. That witness is not a network, so for
physical devices only the lower bound~\eqref{eq:costlower} is asserted here;
the matching physical upper bound comes from the Chebyshev chain below.

\subsection{A chain with exact Chebyshev transmission}
\label{sec:chebyshev}

The upper side of the contact law needs a chain whose transmission ripples
by a prescribed small amount on the window and rises exponentially outside
it. Let $n\ge3$ be odd, $\epsilon,w>0$, $u=\operatorname{asinh}(1/\epsilon)$,
$\theta=\pi/n$ and $\eta=\sinh(u/n)$. The chain with zero diagonal and
\begin{equation}
t_j^2=\frac{w^2\big[\eta^2+\sin^2(j\theta)\big]}{4\sin\big((j-\tfrac12)\theta\big)\sin\big((j+\tfrac12)\theta\big)}\quad(1\le j<n),\qquad
\Gamma=\frac{w\eta}{2\sin(\theta/2)},
\label{eq:cheb}
\end{equation}
transmits, for every real $E$,
\begin{equation}
T(E)=\frac1{1+\epsilon^2\mathsf T_n(E/w)^2}.
\label{eq:chebT}
\end{equation}

The coefficients~\eqref{eq:cheb} are the equal-contact form of the classical
Chebyshev filter prototype~\cite{Cohn_1957dc}; the proof below establishes the
exact transmission with Markovian contacts at every odd order, without a
narrow-band approximation. Every hopping is positive because the sines in the
denominator have arguments in $(0,\pi)$. The transmission ripples between
$1/(1+\epsilon^2)$ and $1$ on the window $|E|\le w$, and $\mathsf T_n(E/w)=\cosh(n\operatorname{acosh}(E/w))$ grows exponentially outside it.

\textit{Similarity to a solvable operator.} Take $w=1$ and put $a=u/n$. Index the sites by
$j=0,\dots,n-1$, and let $y_j=(j+\tfrac12)\theta$ and $b=\theta/2+ia$.
Define
\begin{equation}
A_j=\frac{\sin(y_j+b)}{2\sin y_j},\qquad B_j=\frac{\sin(y_j-b)}{2\sin y_j},\qquad
(\mathsf Lv)_j=A_jv_{j+1}+B_jv_{j-1},
\end{equation}
with reflecting ghost values $v_{-1}=v_0$ and $v_n=v_{n-1}$. The interior
diagonal of $\mathsf L$ vanishes, and its end entries are
$B_0=\sin(-ia)/(2\sin(\theta/2))=-i\Gamma$ and
$A_{n-1}=\sin(\pi+ia)/(2\sin(\theta/2))=-i\Gamma$. With
$|\sin(x+iy)|^2=\sin^2x+\sinh^2y$,
\begin{equation}
A_{j-1}B_j=\frac{\sin(j\theta+ia)\sin(j\theta-ia)}{4\sin((j-\tfrac12)\theta)\sin((j+\tfrac12)\theta)}=t_j^2 .
\end{equation}
No $A_j$ with $j<n-1$ vanishes, so a diagonal similarity
$\mathsf D^{-1}\mathsf L\mathsf D$ with $d_{j+1}/d_j=t_{j+1}/A_j$ produces
equal off-diagonal entries $t_j$ and leaves the diagonal unchanged. Hence
$\mathsf L$ is similar to $\heff=h-i\Gamma(\Pi_1+\Pi_n)$ and has the same
characteristic polynomial.

The spectrum follows from a polynomial flag. A polynomial in $\cos y$ takes
equal values at $y=\pm\theta/2$ and at $y=\pi\pm\theta/2$, so the ghost
prescription is its ordinary continuation, and $\mathsf L$ acts on node values
of $f(y)$ as $A(y)f(y+\theta)+B(y)f(y-\theta)$. For $f_k(y)=\cos(ky)=\mathsf T_k(\cos
y)$, addition formulas give
\begin{equation}
(\mathsf Lf_k)(y)=\cos b\cos(k\theta)\cos(ky)-\sin b\sin(k\theta)\cot y\sin(ky).
\end{equation}
For $k\ge1$, $\cot y\sin(ky)=xU_{k-1}(x)$ with $x=\cos y$ and $U_{k-1}$ the
Chebyshev polynomial of the second kind; it has the same degree $k$ and
leading coefficient $2^{k-1}$ as $\mathsf T_k$. Hence $\mathsf
L\mathsf T_k=\cos(b+k\theta)\mathsf T_k+(\text{degree}<k)$, and $\mathsf L1=\cos b$ since
$A_j+B_j=\cos b$. The nodes $\cos y_j$ are distinct, so the evaluated
$\mathsf T_0,\dots,\mathsf T_{n-1}$ form a basis of $\mathbb C^n$ in which $\mathsf L$ is
triangular. The eigenvalues of $\heff$ are therefore
\begin{equation}
z_k=\cos\big((k+\tfrac12)\theta+ia\big),\qquad k=0,\dots,n-1,
\label{eq:chebpoles}
\end{equation}
distinct, with $\operatorname{Im}z_k=-\sinh a\sin((k+\tfrac12)\theta)<0$.
At each pole $\mathsf T_n(z_k)=\cos((k+\tfrac12)\pi+iu)=(-1)^{k+1}i\sinh u$, so the
$n$ poles and their conjugates are the $2n$ roots of
$\mathsf T_n^2+\sinh^2u$, and matching leading coefficients,
\begin{equation}
|D(E)|^2=2^{2-2n}\big(\mathsf T_n(E)^2+\sinh^2u\big),\qquad D(E)=\det(E-\heff).
\end{equation}
For the numerator of Eq.~\eqref{eq:cofactor}, the product
$\prod_{j=0}^{n-1}\sin(z+j\pi/n)=2^{1-n}\sin(nz)$, which follows from the
factorization of $\zeta^n-1$, gives at $z=ia$ and at $z=\theta/2$
\begin{equation}
\prod_{j=1}^{n-1}\big(\sinh^2a+\sin^2(j\theta)\big)=2^{2-2n}\frac{\sinh^2u}{\sinh^2a},\qquad
\prod_{j=1}^{n-1}4\sin\big((j-\tfrac12)\theta\big)\sin\big((j+\tfrac12)\theta\big)=\frac1{\sin^2(\theta/2)} .
\end{equation}
With $4\Gamma^2=\sinh^2a/\sin^2(\theta/2)$ these give
$4\Gamma^2\tau^2=2^{2-2n}\sinh^2u$. Hence
$T=\sinh^2u/(\mathsf T_n^2+\sinh^2u)=1/(1+\epsilon^2\mathsf T_n^2)$, since $\sinh u=1/\epsilon$.
Restoring $w$ multiplies both sides of each identity by $w^{2n}$, which gives
Eq.~\eqref{eq:chebT}.

\subsection{Resource scaling and the contact law}
\label{sec:chebfamily}

Tuned to a target product, the Chebyshev chain reaches it at a contact
cost that exceeds the lower bound only by an additive constant, while its
current stays near $w/\pi$. For $0<p\le1$ and any $w>0$ prescribe
\begin{equation}
S_0=\log\frac{16}p,\quad A=S_0+\log S_0,\quad \epsilon^2=\frac p{16A},\quad
\ell_\epsilon=\log\frac2\epsilon,
\label{eq:chebfamily}
\end{equation}
and let $n$ be the smallest odd integer with
$n\ge\max\{3,\ell_\epsilon,[256A(\ell_\epsilon+1)/p]^{1/2}\}$. The Chebyshev
chain of Sec.~\ref{sec:chebyshev} with these $n$, $\epsilon$, $w$ at
affinity $A$ has $P<p$, and
\begin{equation}
\frac{wq}{\pi(1+\epsilon^2)}\le j\le\frac{wq}\pi\Big[1+\frac{2(\ell_\epsilon+1)^2}{n^2}\Big].
\label{eq:chebcurrent}
\end{equation}
As $p\downarrow0$, with $s=\log(1/p)$, $n\sim8\sqrt2\,s/\sqrt p$,
$j=(w/\pi)[1+O(p)]$ and
\begin{equation}
\frac{\Gamma}{j}=\frac s2+\frac{\log s}2+\log8+o(1).
\label{eq:chebcost}
\end{equation}
The derivation first bounds the transmission integrals of any Chebyshev
chain, then inserts the prescription.

\textit{Window integrals.} Let $0<\epsilon\le1$ and $n\ge\max(3,\ell_\epsilon)$,
and put $J_k=\pi I_k/w$. Outside the window substitute
$E/w=\cosh(s'/n)$, so $\mathsf T_n=\cosh s'$; the out-of-band mass and noise are
\begin{equation}
M_{\rm o}=\int_0^\infty\frac{\sinh(s'/n)}n\mathcal Q(s')\,ds',\qquad
N_{\rm o}=\int_0^\infty\frac{\sinh(s'/n)}n\mathcal Q(s')\big(1-\mathcal Q(s')\big)\,ds',\qquad
\mathcal Q=\frac1{1+\epsilon^2\cosh^2s'},
\end{equation}
and evenness gives $J_1=\int_0^1T(wx)\,dx+M_{\rm o}$ and
$J_2=\int_0^1[T(1-T)](wx)\,dx+N_{\rm o}$. From $e^{s'}/2\le\cosh s'\le e^{s'}$ and
$e^{-\ell_\epsilon}=\epsilon/2$, $z=\epsilon^2\cosh^2s'$ lies between
$e^{2(s'-\ell_\epsilon)}$ and $4e^{2(s'-\ell_\epsilon)}$, so
$\mathcal Q\le\min(1,e^{-2(s'-\ell_\epsilon)})$ and
$\mathcal Q(1-\mathcal Q)\le4e^{-2|s'-\ell_\epsilon|}$. With
$\sinh(s'/n)/n\le s'e^{s'/n}/n^2$ and $e^{s'/n}\le e\,e^{|s'-\ell_\epsilon|/3}$,
splitting at $s'=\ell_\epsilon$ and using $\cosh y-1\le ey^2/2$ for $y\le1$,
\begin{equation}
M_{\rm o}\le\frac e{n^2}\Big(\frac{\ell_\epsilon^2}2+\frac{3\ell_\epsilon}5+\frac9{25}\Big)\le\frac{2(\ell_\epsilon+1)^2}{n^2},\qquad
N_{\rm o}\le\frac{4e}{n^2}\int_{\mathbb R}(\ell_\epsilon+|v|)e^{-5|v|/3}dv\le\frac{16(\ell_\epsilon+1)}{n^2}.
\label{eq:chebtails}
\end{equation}
The leakage beyond the window costs noise of order $\ell_\epsilon/n^2$ and mass
of order $\ell_\epsilon^2/n^2$; neither is dropped. Inside the window
$|\mathsf T_n|\le1$ and $\int_0^1\mathsf T_n^2dx=\tfrac12-1/[2(4n^2-1)]\le\tfrac12$, by the
substitution $x=\cos\phi$. From $1-z\le(1+z)^{-1}\le1$ and
$z/(1+z)^2\le z$ with $z=\epsilon^2\mathsf T_n^2$,
$1-\epsilon^2/2\le J_1\le1+M_{\rm o}$ and $J_2\le\epsilon^2/2+N_{\rm o}$, hence
\begin{equation}
\frac w{\pi(1+\epsilon^2)}\le I_1\le\frac w\pi\Big[1+\frac{2(\ell_\epsilon+1)^2}{n^2}\Big],\qquad
F_0\le(1+\epsilon^2)\Big[\epsilon^2+\frac{16(\ell_\epsilon+1)}{n^2}\Big].
\label{eq:chebI}
\end{equation}

\textit{The prescription.} The prescription has $\epsilon^2\le1$ and
$16(\ell_\epsilon+1)/n^2\le p/(16A)=\epsilon^2$, so
$F_0\le2\epsilon^2(1+\epsilon^2)\le p/(4A)$, and the partition term of
Eq.~\eqref{eq:Pformula} is at most $p/2$. Since $S_0>2$, $\log S_0\le S_0/2$
and $e^{-A}=p/(16S_0)\le1/16$,
$\beta(A)=4Ae^{-A}/(1-e^{-2A})\le(3p/8)/(1-1/256)<p/2$. Hence $P<p$, and
Eq.~\eqref{eq:chebcurrent} is $j=qI_1$ with Eq.~\eqref{eq:chebI}. As
$p\downarrow0$, $A=s+\log s+\log16+o(1)$, $\epsilon\sim\tfrac14\sqrt{p/s}$,
$\ell_\epsilon=s/2+\tfrac12\log s+\log8+o(1)$, $u=\ell_\epsilon+O(\epsilon^2)$,
and $n\sim(256A\ell_\epsilon/p)^{1/2}\sim8\sqrt2\,s/\sqrt p$; rounding to the
next odd integer changes none of these limits. Then $q=1-O(p/s)$ and
$J_1=1+O(p)$, so $j=(w/\pi)[1+O(p)]$. Since $u/n=O(\sqrt p)$,
$\Gamma=w\sinh(u/n)/[2\sin(\pi/2n)]=(wu/\pi)[1+O(p)]$, and
$\Gamma/j=u+O(pu)$, which is Eq.~\eqref{eq:chebcost}.

The coefficients of the family follow from Eq.~\eqref{eq:cheb}, rewritten as
$t_j^2=\tfrac{w^2}4\{1+[\eta^2+\sin^2(\theta/2)]/[\sin^2(j\theta)-\sin^2(\theta/2)]\}$.
The largest hopping is at the ends and the smallest in the middle, so
$t_{\min}=\tfrac w2[1+O(p)]$, $t_{\max}=w\sqrt{u^2+\pi^2}/(\pi\sqrt3)\,[1+O(p)]$
and $\Gamma/t_{\min}\sim2u/\pi$. The contact-to-bulk ratio grows as $\log(1/p)$,
and no common rescaling changes it.

The two sides together give the contact law of the main text. We call
$\mathcal C(p)$ the infimum of $\Gamma_*/j$ over all $n$, all Hermitian
$h$, all $\Gamma_{\rm L},\Gamma_{\rm R}>0$ with $K$ strictly stable and
$I_1>0$, and all finite $A>0$, subject to $P\le p$. Then, as $p\downarrow0$,
\begin{equation}
\mathcal C(p)=\frac12\log\frac1p+\frac12\log\log\frac1p+O(1).
\label{sm-eq:contactlaw}
\end{equation}
The Chebyshev chain is a member of this class with
$\Gamma_*=\Gamma$, so Eq.~\eqref{eq:chebcost} bounds $\mathcal C(p)$ from
above. Equations~\eqref{eq:costlower} and \eqref{eq:squeeze} bound it from below by
$s/2+\tfrac12\log s+\tfrac12\log2$.

The centered infimum thus lies between $\tfrac12\log2$ and $\log8$ in the limit;
the additive constant and attainment are not decided. The law holds over all
network sizes, so it is not a fixed-size statement. Bounded $\Gamma_*$ with
$P\to0$ forces $j\to0$. Here $\Gamma_*$ is the single-particle linewidth of the
weaker contact, not its stationary jump activity.

\subsection{The trigonometric chain}

For comparison, the trigonometric chain of Ref.~\cite{Brandner_2025tu} at
half-bandwidth $w$ has $T=1/[1+(E/w)^{2n}]$, $\Gamma_{\rm
tr}=w/[2\sin(\pi/2n)]$, $I_1=\Gamma_{\rm tr}/n$ and $F_0=1/(2n)$, by
Sec.~\ref{sec:transmission} and the common scaling. At finite affinity,
\begin{equation}
P_{\rm tr}=\frac{Aq}n+\beta(A),\qquad\frac{\Gamma_{\rm tr}}{j_{\rm tr}}=\frac nq .
\label{eq:trig}
\end{equation}
The partition noise falls only as $1/n$, so the length, and with it the
contact cost, must grow as $1/p$. For $s\ge8$, $P_{\rm tr}\le p$ forces
$A>s$ as in the lower half of Eq.~\eqref{eq:squeeze}, and $n\ge Aq/(p-\beta(A))$,
so $\Gamma_{\rm tr}/j_{\rm tr}\ge A/(p-\beta(A))>s/p$. Conversely, at $A_t$
and $n=\lceil A_tq_t/(p-\beta(A_t))\rceil$,
$\Gamma_{\rm tr}/j_{\rm tr}\le A_t/[p(1-2/s)]+2$. The cost is therefore
$\Theta(\log(1/p)/p)$, against $\Theta(\log(1/p))$ for the Chebyshev chain.
More precisely, the continuous relaxation $A/(p-\beta(A))$ is minimal where
$p=\beta-A\beta'=2A^2\cosh A/\sinh^2A$. For small $p$ every feasible affinity
exceeds $a_p>2$, and there the right side decreases,
because its logarithmic derivative $2/A+\tanh A-2\coth A$ is below $2/A-1$.
This gives $A_*=s+2\log s+\log4+o(1)$ and $\beta(A_*)/p=\tanh A_*/A_*$, and
rounding $n$ up adds at most $1/q\le2$, so
\begin{equation}
\inf_{n,A:\,P_{\rm tr}\le p}\frac{\Gamma_{\rm tr}}{j_{\rm tr}}=\frac{s+2\log s+\log4+1+o(1)}{p}.
\label{eq:trigcost}
\end{equation}

\section{Vanishing product at bounded couplings}
\label{sec:bounded}

This section proves the bounded-coupling result of the main text. Its first
half is a floor: at a bounded contact linewidth, a product tending to zero
forces the current to zero. Its second half is a construction. Chains whose hoppings all lie in $[1,6]$ and whose two linewidths both equal $1$ reach every target chain with zero on-site energies and equal contacts, scaled down in energy. The error is small on the whole energy axis relative to the transmitted weight. Applied to Chebyshev targets
of growing order, this drives the product to zero.

\subsection{A floor at finite current}
\label{sec:floor}

At finite current through a bounded contact the budget leaves a floor on the
product. Every two-contact network satisfying the hypotheses of
Eq.~\eqref{eq:poleszeros}, at every finite affinity $A>0$, obeys
\begin{equation}
P\ge2\tanh\tfrac12\,\exp\!\Big(-\frac{2\Gamma_*}{j}\Big).
\label{sm-eq:floor}
\end{equation}
Indeed, by Eq.~\eqref{sm-eq:budget},
$\log(1/F_0)\le2\Gamma_*(1-F_0)/I_1\le2\Gamma_*/j$, since $j=qI_1\le I_1$. Hence
$F_0\ge e^{-2\Gamma_*/j}$. In Eq.~\eqref{eq:Pformula}, for $A\ge1$ the
partition term is $2A\tanh(A/2)F_0\ge2\tanh(\tfrac12)F_0$; for $A<1$ the
thermal term is $2A/\sinh A>2/\sinh1>2\tanh\tfrac12\ge2\tanh(\tfrac12)F_0$.

Along a sequence with $\Gamma_*\le R$ and $P\to0$ the right side of
Eq.~\eqref{sm-eq:floor} must vanish, so $j\to0$. The floor uses no structure of
the network beyond the budget.

\subsection{The embedding construction}

Fix a target: an $n$-site chain, $n\ge2$, with zero on-site energies,
hoppings $t_j^*>0$, both linewidths $\Gamma^*>0$, and transmission $T^*(x)$.
Its moments $I_1^*=\int T^*dx/2\pi>0$ and
$F_0^*=\int T^*(1-T^*)dx/(2\pi I_1^*)$ are finite. Put
\begin{equation}
\zeta_1=\zeta_n=\frac{323}{280},\qquad \zeta_j=\frac{37}{35}\ (1<j<n),\qquad
Z=\operatorname{diag}(\zeta_1,\dots,\zeta_n).
\label{eq:zeta}
\end{equation}
For an integer depth $\nu$ define
\begin{equation}
\tau_\nu=9^{-\nu},\quad \delta_\nu=\frac{\tau_\nu}{\zeta_1\Gamma^*},\quad
\alpha_{j}=\delta_\nu t_j^*\sqrt{\zeta_j\zeta_{j+1}},\quad
\ell_\nu=\Big\lceil\max_j\frac{\log(1/\alpha_j)}{\log6}\Big\rceil,\quad
v_j=\alpha_j^{-1/\ell_\nu}.
\label{eq:embedparams}
\end{equation}
For $\nu$ large enough $0<\alpha_j<1$ and $\ell_\nu\ge1$, so $1<v_j\le6$. The
physical chain has zero on-site energies, both linewidths equal to $1$, and
the bonds, read from left to right,
\begin{equation}
(3,1)^\nu\;\prod_{j=1}^{n-1}\big[(1,v_j)^{\ell_\nu}\,1\big]\;(1,3)^\nu ,
\label{eq:bondword}
\end{equation}
where the product is concatenation. Every bond lies in $[1,6]$, and the chain
has $m(n,\nu)=4\nu+(n-1)(2\ell_\nu+1)+1$ modes. With
$\vartheta=\log9/\log6$, writing $\log(1/\alpha_j)=\nu\log9+O_*(1)$ gives
\begin{equation}
\ell_\nu=\vartheta\nu+O_*(1),\qquad v_j=6+O_*(\nu^{-1}).
\label{eq:ellnu}
\end{equation}
A star on $O$ or $C$ marks dependence on the fixed target; no constant is
claimed uniform in the target. In particular $v_j\ge23/4$ for $\nu$ large.

The mechanism is visible at zero energy. A pair of a strong bond $v$ and a weak bond $1$ passes an amplitude $1/v$, so $\ell$ such pairs act as one weak hopping $v^{-\ell}$. The arms $(3,1)^\nu$ likewise weaken the unit contact to a linewidth $\tau_\nu$ at the first retained site. The retained sites are
the first site after the left arm and the ends of successive central words,
$n$ in all. The weights $\zeta_j$ are their inverse quasiparticle weights. Effective weak hoppings from dimerized segments and their decimation are used for transfer across topological domain walls~\cite{Zurita_2023fq,Palavicini_2026hq}. What is added here is the constant Markovian contact and the error control on the whole energy axis.

The embedded chain reproduces the scaled target transmission on the whole
energy axis, with an error small against the transmitted weight. Let $T_\nu$
be the transmission of the chain~\eqref{eq:bondword}. There are
target-dependent $C_*$, $\nu_*$ and a target-independent $\lambda<1$ such
that, for $\nu\ge\nu_*$,
\begin{equation}
\frac{\mathcal E_\nu}{\delta_\nu}\le
C_*\Big[\nu^{-3/8}+\delta_\nu\nu^{3/8}+(\nu+1)^2\lambda^\nu\Big],\qquad
\mathcal E_\nu=\int_{\mathbb R}\big|T_\nu(E)-T^*(E/\delta_\nu)\big|\frac{dE}{2\pi}.
\label{eq:embederror}
\end{equation}

Pointwise convergence of $T_\nu(\delta_\nu x)$ to $T^*(x)$ on compact sets would not suffice. The product needs the error small against the transmitted weight $\delta_\nu I_1^*$, which shrinks, and narrow resonances far outside the window could carry weight comparable to it. The derivation below controls them
through the spectral mass of the end sites.

\textit{Exact blocks.} For the closed $2s$-site block $\mathsf B_{v,s}$ with
bonds $(v,1)^{s-1}v$ put $x_v(E)=(E^2-v^2-1)/(2v)$ and
$\mathcal P_{v,s}=U_s(x_v)+v^{-1}U_{s-1}(x_v)$, with $U_s$ the Chebyshev
polynomial of the second kind and $U_{-1}=0$. The continuant recurrence, with
initial values $1$, $E$, $E^2-v^2$, gives
\begin{equation}
\det(E-\mathsf B_{v,s})=v^s\mathcal P_{v,s},\qquad
b_{v,s}\equiv(E-\mathsf B_{v,s})^{-1}_{1,2s}=\frac1{\mathcal P_{v,s}},\qquad
d_{v,s}\equiv(E-\mathsf B_{v,s})^{-1}_{11}=\frac Ev\frac{U_{s-1}(x_v)}{\mathcal P_{v,s}}.
\label{eq:blocks}
\end{equation}
The left arm, with its unit contact at the outer end, gives the first
retained site the self-energy $\sigma_\nu=u_\nu-i\gamma_\nu$ with
\begin{equation}
\gamma_\nu(E)=\Big|U_\nu(x_3)+\frac{1+iE}3U_{\nu-1}(x_3)\Big|^{-2},
\label{eq:gammanu}
\end{equation}
and the right arm, its mirror image, gives the last retained site the same
self-energy. Eliminating the arms and the central blocks, with
$d_j=d_{v_j,\ell_\nu}$ and $b_j=b_{v_j,\ell_\nu}$, leaves the $n\times n$
retained matrix $\mathsf M(E)$: off-diagonal entries $-b_j$, interior
diagonal entries $E-d_{j-1}-d_j$, and end entries $E-\sigma_\nu-d_1$ and
$E-\sigma_\nu-d_{n-1}$. Tracing the two end-to-retained amplitudes gives
\begin{equation}
T_\nu(E)=4\gamma_\nu(E)^2\big|(\mathsf M(E)^{-1})_{n1}\big|^2 .
\label{eq:Tretained}
\end{equation}
At $E=0$ the hyperbolic form of $U_s$ gives $\mathcal P_{v,s}(0)=(-1)^sv^s$,
so $b_j(0)=(-1)^{\ell_\nu}\alpha_j$, and $d_j(0)=0$, $\sigma_\nu(0)=-i\tau_\nu$.
With $S_s(v)=(1-v^{-2s})/(v^2-1)$, the derivatives at zero are
$d_j'(0)=-S_{\ell_\nu}(v_j)$ and $\sigma_\nu'(0)=-S_\nu(3)(1-\tau_\nu)$, from
the arm recurrence $\sigma_0=-i$,
$\sigma_l=(E-\sigma_{l-1})/[E(E-\sigma_{l-1})-9]$. As $S_s(6)\to1/35$ and
$S_\nu(3)\to1/8$, the diagonal of $\mathsf M'(0)$ is $Z+O_*(1/\nu)$, which is
where Eq.~\eqref{eq:zeta} comes from.

\textit{Compact window.} After a diagonal sign change removing
$(-1)^{\ell_\nu}$ from the links,
\begin{equation}
\delta_\nu^{-1}Z^{-1/2}\mathsf M(\delta_\nu x)Z^{-1/2}
=x-h^*+i\Gamma^*(\Pi_1+\Pi_n)+O_*\Big(\frac{|x|}\nu+\delta_\nu x^2\Big),
\label{eq:compactlimit}
\end{equation}
since $\alpha_j/(\delta_\nu\sqrt{\zeta_j\zeta_{j+1}})=t_j^*$ and
$\tau_\nu/(\delta_\nu\zeta_1)=\Gamma^*$. The remainder is uniform on a fixed small disk of energies. Let $g_o$ and $b_o$ be the end diagonal and cross entry of the closed arm. The arm's self-energy has the rank-one form $\sigma_\nu=g_o-ib_o^2/(1+ig_o)$; since $g_o(0)=0$ its denominator is $1$ at zero, so $\sigma_\nu$ is analytic with uniformly bounded derivatives on the disk. The closed outer and central inverses at zero have
norms at most $1/2$ and $1/(v_j-1)\le4/19$, so a Neumann expansion bounds
their derivatives uniformly in $\nu$. Also
$\gamma_\nu(\delta_\nu x)/\tau_\nu=1+O_*(\delta_\nu x^2)$. The target matrix $x-h^*+i\Gamma^*(\Pi_1+\Pi_n)$ is invertible for real $x$: a real eigenvalue would force an eigenvector vanishing at an end, hence everywhere by the recurrence. Its inverse tends to zero at infinity and is therefore bounded on the real axis by a target constant. With Eq.~\eqref{eq:Tretained},
for $M=\nu^{1/8}$ and $\nu$ large,
\begin{equation}
\sup_{|x|\le M}\big|T_\nu(\delta_\nu x)-T^*(x)\big|\le C_*\Big(\frac M\nu+\delta_\nu M^2\Big).
\label{eq:compacterror}
\end{equation}

\textit{Block estimates.} Inside the gap of the dimerized block, $0\le E<v-1$, let
\begin{equation}
\varrho_v(E)=\frac{v^2+1-E^2-\sqrt{(v^2+1-E^2)^2-4v^2}}{2v},
\end{equation}
which increases from $1/v$ to $1$. For $E\ge4$ define $\varrho_3$ by the same formula with
$|10-E^2|$ in place of $10-E^2$, decreasing from $1$. The hyperbolic form of
$U_s$ gives $|\mathcal P_{v,s}|\ge\varrho_v^{-s}$ and
$|U_s|\le(s+1)\varrho_v^{-s}$, hence from Eq.~\eqref{eq:gammanu}
\begin{equation}
\gamma_\nu^{-1}\le C(\nu+1)^2\varrho_3^{-2\nu}\ (1\le E\le2),\qquad
\gamma_\nu^{-1}\le C(\nu+1)^2\ (2\le E\le4),\qquad
\gamma_\nu\le\varrho_3^{2\nu}\ (0\le E\le1\ \text{or}\ E\ge4),
\label{eq:gammabounds}
\end{equation}
where on $[2,4]$, the band of the arm, $|U_s|\le s+1$. Since
$v_j\ge23/4$, every central block stays inside its gap for $E\le9/2$, and
Eq.~\eqref{eq:ellnu} with the smoothness of $\log\varrho_v(E)$ gives
\begin{equation}
\mathsf b_\nu(E)\equiv\max_j|b_j(E)|\le C_*\varrho_6(E)^{\vartheta\nu}\qquad(0\le E\le9/2).
\label{eq:bbound}
\end{equation}
Each block is a disjoint set of strong dimers plus weak bonds of norm at most
$1$, so its spectrum has $|\lambda|\ge v-1$, and bipartite symmetry gives
$d_j(E)=E[(E^2-\mathsf B^2)^{-1}]_{11}<0$ for $0<E\le9/2$. For the arm, the rank-one form gives $\operatorname{Re}\sigma_\nu=g_o[1-b_o^2/(1+g_o^2)]$. On $(0,1]$ the closed arm is bipartite with spectrum outside $(-2,2)$, so $g_o<0$ as for $d_j$, and $|b_o|\le\varrho_3(1)^\nu<1$; hence $\operatorname{Re}\sigma_\nu\le0$. For $E\ge4$ the recurrence $g_0=0$, $g_l=(E-g_{l-1})/[E(E-g_{l-1})-9]$ keeps $0\le g_o\le1$; hence $\operatorname{Re}\sigma_\nu\le1$.

\textit{Spectral domination.} Let $a$ and $b$ be the first and last retained
sites and $\mathcal S(E)=-\operatorname{Im}[G_{aa}(E)+G_{bb}(E)]$, with $G$ the
resolvent of the whole physical chain. All poles of $G$ lie below the real
axis by the recurrence argument above, and $G_{jj}(z)=1/z+O(z^{-2})$, so
closing the contour in the upper half plane gives
\begin{equation}
\int_{\mathbb R}\mathcal S(E)\frac{dE}{2\pi}=1 .
\label{eq:summass}
\end{equation}
Eliminating only the arms, the dissipation identity gives
$-\operatorname{Im}G_{aa}=\gamma_\nu(|G_{aa}|^2+|G_{ab}|^2)$ and its mirror,
whose sum is at least $2\gamma_\nu|G_{ab}|^2$, while
$T_\nu=4\gamma_\nu^2|G_{ab}|^2$. Hence $T_\nu\le2\gamma_\nu\mathcal S$ at every
energy, with every central resonance retained. On $1\le E\le9/2$, eliminate
in addition the internal retained sites, whose real matrix has diagonal at
least $E$ and off-diagonal entries at most $\mathsf b_\nu\to0$. This leaves
\begin{equation}
\begin{pmatrix}y_1+i\gamma_\nu&-\kappa\\-\kappa&y_2+i\gamma_\nu\end{pmatrix},\qquad
y_1,y_2,\kappa\in\mathbb R,\quad |\kappa|\le C_*\mathsf b_\nu,
\end{equation}
with $y_1,y_2\ge E-1-2\mathsf b_\nu^2/E\ge5/2$ on $[4,9/2]$. Direct inversion
gives $T_\nu/\mathcal S=4\gamma_\nu\kappa^2/(y_1^2+y_2^2+2\gamma_\nu^2+2\kappa^2)$,
so altogether
\begin{equation}
T_\nu\le\min\Big(2\gamma_\nu,\frac{2\kappa^2}{\gamma_\nu}\Big)\mathcal S .
\label{eq:domination}
\end{equation}
Arbitrarily narrow resonances are thus charged against the total spectral
mass, with no assumption on their widths or multiplicities.

\textit{Leakage outside the window.} The transmission is even in $E$. On
$[1,2]$ and $[2,4]$, Eqs.~\eqref{eq:summass}--\eqref{eq:domination} with
Eqs.~\eqref{eq:gammabounds} and \eqref{eq:bbound} bound the integrated
transmission by $C_*(\nu+1)^2\lambda_1^\nu$ and $C_*(\nu+1)^2\lambda_2^\nu$,
with $\lambda_1=\sup_{[1,2]}\varrho_6^{2\vartheta}/\varrho_3^2$ and
$\lambda_2=\varrho_6(4)^{2\vartheta}$. On $[4,9/2]$ the real parts
$y_1,y_2\ge5/2$ bound the determinant away from zero, so
$T_\nu\le C\gamma_\nu^2\kappa^2$; splitting at $17/4$ gives the bases
$\lambda_3=\varrho_6(17/4)^{2\vartheta}$ and
$\lambda_4=\varrho_3(17/4)^4\varrho_6(9/2)^{2\vartheta}$. Above $9/2$,
$T_\nu\le2\gamma_\nu\mathcal S$ and Eq.~\eqref{eq:summass} give at most
$2\varrho_3(9/2)^{2\nu}$, base $\lambda_5=\varrho_3(9/2)^2$. Every base is below
$1/9$. For $\lambda_1$, the logarithmic derivative of
$\varrho_6^{2\vartheta}/\varrho_3^2$,
$4E\{\vartheta[(37-E^2)^2-144]^{-1/2}-[(10-E^2)^2-36]^{-1/2}\}$, is negative on
$[1,2)$ because $\vartheta<5/4$, so the supremum sits at $E=1$, where
$\varrho_6(1)=3-\sqrt8<7/40$, $\varrho_3(1)=(3-\sqrt5)/2>3/8$ and
$\vartheta>6/5$ give $\lambda_1<(7/40)^{12/5}(8/3)^2<1/9$. Since
$6^\vartheta=9$, $\varrho_6(4),\varrho_6(17/4)<1/\sqrt6$ give
$\lambda_2,\lambda_3<1/9$; $\varrho_3(17/4),\varrho_6(9/2)<1/2$ give
$\lambda_4<1/9$; and $x_3(9/2)=41/24>5/3$ gives $\varrho_3(9/2)<1/3$. With
$\lambda<1$ the largest ratio $9\lambda_i$,
\begin{equation}
\int_{|E|>1}T_\nu(E)\frac{dE}{2\pi}\le C_*\tau_\nu(\nu+1)^2\lambda^\nu .
\label{eq:highleak}
\end{equation}

\textit{Leakage inside the window, outside the target.} Put
$\eta_\nu=\nu^{-1/2}$. On $0<E\le1$ every real diagonal entry of $\mathsf M$
is at least $E$, since $d_j<0$ and $\operatorname{Re}\sigma_\nu\le0$. Write
$\mathsf M=\mathsf D-\mathsf K$ with $\mathsf D$ diagonal; where
$\|\mathsf K\|\le2\mathsf b_\nu\le E/2$, a Neumann expansion gives
$|(\mathsf M^{-1})_{n1}|\le2\|\mathsf K\|/E^2$, because $\mathsf D^{-1}$ has
no $(n,1)$ entry. On $[0,\eta_\nu]$, analyticity in $E^2$ of
$\log(v\varrho_v)$ and $\log(3\varrho_3)$ gives
$\gamma_\nu,\mathsf b_\nu\le C_*\tau_\nu$, so for $E\ge M\delta_\nu$ with a
target constant in $M$,
\begin{equation}
T_\nu(E)\le C_*\frac{\tau_\nu^4}{E^4},\qquad
\frac1{\delta_\nu}\int_{M\delta_\nu<|E|\le\eta_\nu}T_\nu\frac{dE}{2\pi}\le\frac{C_*}{M^3}.
\label{eq:lowleak}
\end{equation}
On $[\eta_\nu,1]$, $\mathsf b_\nu\le C_*\varrho_6(1)^{\vartheta\nu}$ and
$\gamma_\nu\le\varrho_3(1)^{2\nu}$ give a contribution at most
$C_*\nu^2\varrho_3(1)^{4\nu}\varrho_6(1)^{2\vartheta\nu}$, which is
$o(\tau_\nu)$ because $9\varrho_3(1)^4\varrho_6(1)^{2\vartheta}<9(2/5)^4/5^2<1$.

\textit{The error on the whole axis.} The target transmission is
$1/(1+Q^*)$ with $Q^*$ of degree $2n\ge4$, so
$\int_{|x|>M}T^*dx\le C_*M^{-3}$. Integrating Eq.~\eqref{eq:compacterror}
over $|x|\le M=\nu^{1/8}$ contributes
$C_*(M^2/\nu+\delta_\nu M^3)\le C_*(\nu^{-3/4}+\delta_\nu\nu^{3/8})$ to
$\mathcal E_\nu/\delta_\nu$; the two tails contribute $C_*\nu^{-3/8}$ by
Eq.~\eqref{eq:lowleak}; and Eq.~\eqref{eq:highleak} with the estimate on
$[\eta_\nu,1]$ contributes $C_*(\nu+1)^2\lambda^\nu$ after enlarging
$\lambda<1$. This is Eq.~\eqref{eq:embederror}.

The constants $C_*$ and $\nu_*$ depend on the target through its hoppings,
its linewidth, and the supremum of its real-axis inverse, each finite. The
threshold therefore exists for every fixed target; it is not claimed uniform
in the target order.

\subsection{Chebyshev targets and the vanishing product}
\label{sec:vanishing}

Embedding Chebyshev targets of growing order drives the product to zero
with every coupling bounded. There are chains with zero on-site energies,
every hopping in $[1,6]$, both linewidths equal to $1$, and lengths
$m_k\to\infty$, and finite affinities $A_k=k$, such that
\begin{equation}
P_k\le\frac{4k}{k^2-1}+\frac{2k}{\sinh k}\longrightarrow0 .
\label{eq:Pbounded}
\end{equation}
Their currents tend to zero, as Eq.~\eqref{sm-eq:floor} requires.

An error in transmission small against the transmitted weight changes the
current and the partition noise only by its relative size. Let $T,T'$ be
functions of $E$ with values in $[0,1]$, $\int T'dE/2\pi=\delta J>0$,
$\int T'(1-T')/\int T'=f$, and $\int|T-T'|dE/2\pi\le\epsilon\delta J$ with
$0\le\epsilon<1$. Then the moments $I_1,F_0$ of $T$ obey
\begin{equation}
\Big|\frac{I_1}{\delta J}-1\Big|\le\epsilon,\qquad
F_0\le\frac{f+\epsilon}{1-\epsilon}.
\label{eq:relative}
\end{equation}
The triangle inequality gives $|I_1-\delta J|\le\epsilon\delta J$. Since
$|x(1-x)-y(1-y)|=|x-y||1-x-y|\le|x-y|$ on $[0,1]$, the numerator of $F_0$ is
at most $\delta J(f+\epsilon)$, while $I_1\ge\delta J(1-\epsilon)$.

For Eq.~\eqref{eq:Pbounded}, take as targets the Chebyshev chains of
Sec.~\ref{sec:chebyshev} with $w=1$, odd order $n$ and
$\epsilon=1/n$. They have zero on-site energies, positive hoppings and equal
linewidths, and by Eq.~\eqref{eq:chebI}, since $n\ge\log(2n)$,
\begin{equation}
I_1^*\ge\frac1{\pi(1+n^{-2})},\qquad
F_0^*\le(1+n^{-2})\,\frac{1+16[\log(2n)+1]}{n^2}\longrightarrow0 .
\end{equation}
For each $k\ge2$ choose odd $n_k$ with $F_0^*\le k^{-2}$, and then, by
Eq.~\eqref{eq:embederror}, a depth $\nu_k$ with
$\mathcal E_{\nu_k}\le k^{-2}\delta_{\nu_k}I_1^*$ and
$m(n_k,\nu_k)>m(n_{k-1},\nu_{k-1})$; both conditions hold for every large enough
$\nu$. Equation~\eqref{eq:relative} with $\epsilon=k^{-2}$ gives
$F_{0,k}\le2/(k^2-1)$, and Eq.~\eqref{eq:Pformula} at $A_k=k$ gives
Eq.~\eqref{eq:Pbounded}. Every member is a finite chain with positive
hoppings, so Secs.~\ref{sec:relaxation} and \ref{sec:cumulants} apply: it has a
unique attracting stationary state in the thermal class, and
Eq.~\eqref{eq:Pformula} holds. Since both linewidths equal $1$, Eq.~\eqref{sm-eq:budget} gives
$j_k\le I_{1,k}\le2(1-F_{0,k})/\log(1/F_{0,k})\to0$.

The order of limits matters: for each target the depth is taken large first,
and only then is the target sharpened. At a fixed target order the current falls as $9^{-\nu}$ while the length grows as $[4+2\vartheta(n-1)]\nu$. The current is thus exponentially small in the length, at a rate that falls as $1/n$. Along the sequence, where $n_k$
grows, no rate in the length $m_k$ is claimed.

\section{Every Hilbert-space dimension}
\label{sec:padding}

The chains of Sec.~\ref{sec:sizelaw} have dimension $2^n$. Auxiliary levels
fill in the other dimensions, following the dimension-monotonicity argument of
Ref.~\cite[Sec.~3.1]{Menczel_2021tu}. Fix a base chain with $n\ge2$ modes and
finite $A>0$, with generator $\Lv$, and an integer $k\ge1$. Let $|{\rm
F}\rangle=c_1^\dagger|{\rm vac}\rangle$, so $H_0|{\rm F}\rangle=0$. Add
orthonormal states $|e_1\rangle,\dots,|e_k\rangle$ and set
\begin{equation}
H_N=H\oplus0,\qquad H_{0N}=H_0\oplus I,\qquad
X_N=H_N-H_{0N},\qquad L'=L\oplus0,
\label{eq:padH}
\end{equation}
for both chain jumps $L$, together with one channel per auxiliary level,
\begin{equation}
V_a=|{\rm F}\rangle\langle e_a|,\quad\gamma_a=1,\quad n_a=\delta>0,\quad
\Delta E_a=1,\quad T_a=\frac{1}{\log(1+1/\delta)}.
\label{eq:padV}
\end{equation}
The dimension is $N=2^n+k$. Then $[H_{0N},L']=-AL'$ and $[H_{0N},V_a]=-V_a$,
every occupation, gap and temperature is positive and finite, and the padded
generator $\Lv_\delta$ is the base generator with $H_N$ and $L'$ plus
$\sum_a[(1+\delta)\Dis[V_a]+\delta\Dis[V_a^\dagger]]$. No Hamiltonian term
touches an auxiliary level, and only the right base channel is counted.

\subsection{Padding at small occupation leaves the product unchanged}
\label{sec:padlemma}

The auxiliary levels carry no current, and at small occupation they leave
the counting statistics of the chain unchanged. There is $\delta_0>0$ such
that for $0<\delta<\delta_0$ the padded generator has a unique stationary
state that attracts every density matrix. Each auxiliary current vanishes
exactly, $\sigma_\delta=2Aj_\delta$, and
\begin{equation}
j_\delta\to j,\qquad d_\delta\to d,\qquad P_\delta\to P\qquad(\delta\to0),
\label{eq:padlimit}
\end{equation}
where $j,d,P$ are the values of the base chain.

\textit{The unpadded limit.} At $\delta=0$, write an $N\times N$ matrix in blocks
$Z=\begin{pmatrix}Z_{cc}&Z_{ca}\\Z_{ac}&Z_{aa}\end{pmatrix}$ over the chain and
the auxiliary levels. Absorb the chain rates into jump matrices $F_b$, so the
chain dissipator is $\sum_b\Dis[F_b]$, and put
$R_c=\sum_bF_b^\dagger F_b\succeq0$. Since
$\sum_aV_aZV_a^\dagger=\Tr(Z_{aa})|{\rm F}\rangle\langle{\rm F}|$ and
$\sum_aV_a^\dagger V_a$ is the auxiliary projector, direct substitution gives
\begin{equation}
\begin{gathered}
\Lv_0Z=\begin{pmatrix}\Lv Z_{cc}+\Tr(Z_{aa})|{\rm F}\rangle\langle {\rm F}|&\Phi Z_{ca}\\Z_{ac}\Phi^\dagger&-Z_{aa}\end{pmatrix},\\
\Phi=-iH-\tfrac12R_c-\tfrac12I.
\end{gathered}
\label{eq:anchor}
\end{equation}
Auxiliary populations decay into the anchor state $|{\rm F}\rangle$, and
coherences with the auxiliary levels decay at rate at least $\tfrac12$: if
$\Phi u=\varpi u$, then
$\operatorname{Re}\varpi\,\|u\|^2=-\tfrac12\langle u,R_cu\rangle-\tfrac12\|u\|^2\le-\tfrac12\|u\|^2$.
The map is block triangular in $(Z_{cc},Z_{aa},Z_{ca},Z_{ac})$. Its spectrum,
with multiplicities, is that of $\Lv$, the value $-1$ repeated $k^2$ times, $k$
copies of the spectrum of $\Phi$, and $k$ copies of its conjugate. By
Sec.~\ref{sec:unique}, zero is algebraically simple and every other
eigenvalue has negative real part; the stationary state is $\rho_*\oplus0$.

The entries of $\Lv_\delta$ are affine in $\delta$. The eigenvalues depend
continuously on $\delta$, and zero stays an eigenvalue because $\Lv_\delta$
preserves the trace. For small $\delta$, zero therefore remains algebraically
simple and the rest of the spectrum stays in the open left half-plane. Since
$\Lv_\delta$ generates a completely positive trace-preserving semigroup for
$\delta>0$, its unique stationary state $\rho_\delta$ is a density matrix
attracting every initial state, and $\rho_\delta\to\rho_*\oplus0$.

No Hamiltonian term and no other channel touches $|e_a\rangle$, so for any
state
\begin{equation}
\frac{d}{dt}\langle e_a|\rho|e_a\rangle=-(1+\delta)\langle e_a|\rho|e_a\rangle+\delta\langle {\rm F}|\rho|{\rm F}\rangle=-j_a .
\label{eq:auxcurrent}
\end{equation}
Stationarity gives $j_a=0$ exactly, so the auxiliary channels produce no
entropy although $A_a=\log(1+1/\delta)$ is large:
$\sigma_\delta=A(j_{\rm L}+j_{\rm R})$. The operator
$Q_N=(\sum_j\hat n_j)\oplus I$ commutes with $H_N$ and with every $V_a$,
because $|{\rm F}\rangle$ and each $|e_a\rangle$ carry the value one, so its
stationary balance gives $j_{\rm L}=j_{\rm R}$ and $\sigma_\delta=2Aj_\delta$.

Tilting the right channel does not change the coherence blocks or the
auxiliary block, so at $\delta=0$ the tilted generator has the block
form~\eqref{eq:anchor} with $\Lv$ replaced by its tilt. The characteristic
polynomial of the tilted padded generator is jointly analytic in
$(\delta,s)$, and its root at $(0,0)$ is simple. The implicit-function theorem
gives a root $\lambda(\delta,s)$ jointly analytic near $(0,0)$, with
$\lambda(\delta,0)=0$ and $\lambda(0,s)$ equal to the chain's. For fixed small
$\delta>0$ the other roots stay to its left, so $\lambda(\delta,s)$ is the
scaled cumulant generating function of the padded device. Its derivatives in
$s$ give $j_\delta\to j$ and $d_\delta\to d$; hence $P_\delta=2Ad_\delta/j_\delta\to P$,
which is Eq.~\eqref{eq:padlimit}.

\subsection{The size-law product in every dimension}
\label{sec:dimension}

Padding the size-law chain reaches every Hilbert-space dimension at no
extra entropy production. For every integer $N\ge4$ and $w>0$ let $n=\lfloor\log_2N\rfloor\ge2$ and
$k=N-2^n$. There is an $N$-dimensional member of the thermal-jump class with
positive finite temperatures, a unique attracting stationary state, and
\begin{equation}
P_N<2V_n<66\frac{\log n}{n^2},\qquad j_N>\frac w{4\pi},\qquad \sigma_N=2A_nj_N .
\label{sm-eq:dimension}
\end{equation}
Consequently $0\le\Pmin^N=O(\log\log N/(\log N)^2)$ along all integers
$N\to\infty$.

To build the device, take the size-law chain with $n$ modes at $A_n=2\log n$. If
$k=0$, it is the device, with $P_N\le V_n<2V_n$ and $j_N\ge w/(2\pi)$. If
$k\ge1$, pad it with $k$ levels. By Sec.~\ref{sec:padlemma} choose $\delta>0$ small
enough for unique attraction, $|P_\delta-P_n|<V_n$ and
$|j_\delta-j_n|<w/(4\pi)$; finitely many positive thresholds have a positive
minimum. Then $P_N<P_n+V_n\le2V_n$ and $j_N>j_n-w/(4\pi)\ge w/(4\pi)$, strict
even if the base bounds hold with equality. Finally $\log_2N-1<n\le\log_2N$,
so $\log n=\log\log N+O(1)$ and $n^{-2}=O((\log N)^{-2})$, and
Eq.~\eqref{eq:nonneg} gives $\Pmin^N\ge0$.

The decomposition covers every integer: $N=4,5,8,16,17$ correspond to
$(n,k)=(2,0),(2,1),(3,0),(4,0),(4,1)$. The occupation $\delta$ may depend on
$N$ and $w$; we prove no uniform bound on it, on the auxiliary temperatures,
or on relaxation. The padded devices are not chains, and the lower bound of
Sec.~\ref{sec:lower} does not apply to them.

\section{Transmission of the trigonometric chain at every length}
\label{sec:transmission}

This section proves that the trigonometric chain of
Ref.~\cite{Brandner_2025tu} at half-bandwidth one, with
$\Gamma=1/(2s_0)$ and $t_j=1/(2\sqrt{s_{j-1}s_j})$, where
$s_r=\sin((2r+1)\theta)$ and $\theta=\pi/2n$, transmits
$T(E)=1/(1+E^{2n})$ at every length $n\ge2$, with $I_1=\Gamma/n$ and
$I_2=\Gamma/(2n^2)$. It also gives the roots of $\chi_n(E)=\det(E-\heff)$,
$\heff=h-i\Gamma(\Pi_1+\Pi_n)$. Each $s_r$ is positive and $s_{n-1-r}=s_r$.

\textit{Normalization.} Evaluating
$z^n+1=\prod_{r}(z-e^{i(2r+1)\pi/n})$ at $z=1$ and taking moduli, with
$|1-e^{i\phi}|=2\sin(\phi/2)$, gives
\begin{equation}
\prod_{r=0}^{n-1}s_r=2^{1-n}.
\label{eq:prodsin}
\end{equation}
With $\mathsf D=\operatorname{diag}(\sqrt{2s_0},\dots,\sqrt{2s_{n-1}})$, the
matrix $\mathsf D(E-\heff)\mathsf D$ has off-diagonal entries $-1$, interior
diagonal entries $2Es_r$, and end diagonal entries $2Es_0+i$, because
$2s_0\Gamma=1$. By Eq.~\eqref{eq:prodsin}, $\det\mathsf D^2=2$, so
\begin{equation}
\widetilde\chi_n\equiv\det[\mathsf D(E-\heff)\mathsf D]=2\chi_n.
\label{eq:scaled}
\end{equation}
We show that
\begin{equation}
\chi_n(E)=\prod_{r=1}^{n}\big(E-e^{-i(2r-1)\theta}\big).
\label{eq:roots}
\end{equation}

\textit{A matching identity.} Consider a path of $\ell$ consecutive vertices
$0,\dots,\ell-1$ with vertex weights $s_{k+r}$. Let $S(k,\ell,p)$ be the sum,
over all sets of $p$ disjoint edges, of the product of the weights of the
uncovered vertices ($S=1$ for the empty product). Here $\theta$ is arbitrary.
For $0\le p\le\lfloor\ell/2\rfloor$,
\begin{equation}
\begin{gathered}
S(k,\ell,p)=\Big(\prod_{r=p}^{\ell-1-p}s_{k+r}\Big)g(\ell,p),\\
g(\ell,p)=\prod_{r=1}^{p}\frac{\sin(2(\ell-p-r+1)\theta)}{\sin(2r\theta)},
\end{gathered}
\label{eq:matching}
\end{equation}
wherever the denominators are nonzero. For $p=0$ and for $\ell=2p$ (one
perfect matching, $g(2p,p)=1$) this is immediate. Otherwise, splitting on
whether the last vertex is covered gives
$S(k,\ell,p)=s_{k+\ell-1}S(k,\ell-1,p)+S(k,\ell-2,p-1)$. Inserting
Eq.~\eqref{eq:matching} for the two shorter paths and dividing by the common
factor, the induction step reduces to
\begin{equation*}
\begin{split}
s_{k+\ell-1-p}\sin(2(\ell-p)\theta)={}&s_{k+\ell-1}\sin(2(\ell-2p)\theta)\\
&+s_{k+p-1}\sin(2p\theta),
\end{split}
\end{equation*}
which is the identity
\begin{equation*}
\sin x\sin(u+v)=\sin(x+u)\sin v+\sin(x-v)\sin u
\end{equation*}
at $x=(2(k+\ell-1-p)+1)\theta$, $u=2p\theta$, $v=2(\ell-2p)\theta$. The
division is legitimate on a dense set of $\theta$ and extends by continuity.
In every use below $\ell\le n$ and $p\le n/2$, so $0<2r\theta\le\pi/2$ and all
denominators are positive.

\textit{Expansion at the ends.} Let $F(k,\ell)$ be the determinant of the
$\ell\times\ell$ tridiagonal matrix with off-diagonal entries $-1$ and
diagonal entries $2Es_{k},\dots,2Es_{k+\ell-1}$, with $F(k,0)=1$. Expanding
over matchings, $F(k,\ell)=\sum_p(-1)^p(2E)^{\ell-2p}S(k,\ell,p)$. Expanding
$\widetilde\chi_n$ in the two end entries $i$ and using the reflection
$s_{n-1-r}=s_r$, which makes the two $(n-1)$-dimensional principal minors
equal,
\begin{equation}
\widetilde\chi_n=F(0,n)+2iF(1,n-1)-F(1,n-2).
\label{eq:endexp}
\end{equation}
Writing $\widetilde\chi_n=\sum_{j=0}^n i^j\kappa_jE^{n-j}$,
Eq.~\eqref{eq:endexp} gives $\kappa_0=2^n\prod_rs_r=2$, and
\begin{equation}
\begin{gathered}
\kappa_{2p}=2^{n-2p}\big[S(0,n,p)+S(1,n-2,p-1)\big],\\
\kappa_{2p+1}=2^{n-2p}S(1,n-1,p).
\end{gathered}
\label{eq:kappa}
\end{equation}

\textit{Closed form of the coefficients.} Put
$u_p=\prod_{r=p}^{n-1-p}s_r$, $o_p=\prod_{r=0}^{p-1}s_r$, and
$g_p=\prod_{v=p+1}^{2p-1}\sin(2v\theta)/\prod_{r=1}^{p-1}\sin(2r\theta)$. The
reflection $\sin(2(n-v)\theta)=\sin(2v\theta)$ turns Eq.~\eqref{eq:matching}
into $g(n,p)=g(n-2,p-1)=g_p$, and both matchings in $\kappa_{2p}$ leave the
vertices $p,\dots,n-1-p$ uncovered, so $\kappa_{2p}=2^{n-2p+1}u_pg_p$; this
includes $n=2p$, where $\kappa_n=2$. By reflection and
Eq.~\eqref{eq:prodsin}, $o_p^2u_p=2^{1-n}$, so
$\kappa_{2p}=2^{2-2p}g_p/o_p^2$. Define
\begin{equation}
a_j=\frac{\prod_{v=0}^{j-1}\cos(v\theta)}{\prod_{u=1}^{j}\sin(u\theta)},\qquad a_0=1.
\label{eq:aj}
\end{equation}
Using $\cos(v\theta)=\sin(2v\theta)/(2\sin(v\theta))$ for $1\le v\le2p-1$ and
separating odd and even indices gives $a_{2p}=g_p/(2^{2p-1}o_p^2)$, so
$\kappa_{2p}=2a_{2p}$, and $\kappa_0=2a_0$. For odd orders,
Eq.~\eqref{eq:matching} gives
$S(1,n-1,p)=(u_p/s_p)\prod_{v=p+1}^{2p}\sin(2v\theta)/\prod_{r=1}^p\sin(2r\theta)$,
hence
\begin{equation*}
\frac{\kappa_{2p+1}}{\kappa_{2p}}=\frac{\sin(4p\theta)}{2s_p\sin(2p\theta)}=\frac{\cos(2p\theta)}{\sin((2p+1)\theta)}=\frac{a_{2p+1}}{a_{2p}},
\end{equation*}
and $\kappa_1=2^n\prod_rs_r/s_0=2/s_0=2a_1$. Therefore
$\widetilde\chi_n=2\sum_{j=0}^n i^ja_jE^{n-j}$.

\textit{The roots.} With $\zeta=e^{-i\theta}$, the right side of
Eq.~\eqref{eq:roots} is $\prod_{r=0}^{n-1}(E-\zeta^{2r+1})$. Its coefficient
of $E^{n-j}$ is $(-\zeta)^j$ times the elementary symmetric polynomial of
$1,\zeta^2,\dots,\zeta^{2(n-1)}$,
\begin{equation*}
\mathfrak e_j=\zeta^{j(j-1)}\prod_{u=1}^j\frac{1-\zeta^{2(n-u+1)}}{1-\zeta^{2u}},
\end{equation*}
which follows by induction on the number $\nu$ of variables, at fixed
$\zeta$, from $\mathfrak e_j(\nu)=\mathfrak e_j(\nu-1)+\zeta^{2(\nu-1)}\mathfrak
e_{j-1}(\nu-1)$ and
$(1-\zeta^{2(\nu-j)})+\zeta^{2(\nu-j)}(1-\zeta^{2j})=1-\zeta^{2\nu}$. Since
$n\theta=\pi/2$,
$(1-\zeta^{2(n-u+1)})/(1-\zeta^{2u})=e^{-i(n+1-2u)\theta}\cos((u-1)\theta)/\sin(u\theta)$,
and the phases combine to $(-1)^je^{-ijn\theta}=i^j$. The coefficient is
$i^ja_j$, which proves Eq.~\eqref{eq:roots} through Eq.~\eqref{eq:scaled}.

\textit{Transmission.} Every root in Eq.~\eqref{eq:roots} lies strictly in
the lower half-plane. These $n$ roots and their conjugates are exactly the
$2n$ roots of $E^{2n}+1$, so $|\chi_n(E)|^2=1+E^{2n}$ for real $E$. By
Eq.~\eqref{eq:prodsin} and $s_{n-1}=s_0$,
$\prod_jt_j^2=s_0^2/(4^{n-1}\prod_rs_r^2)=s_0^2$, so $4\Gamma^2\prod_jt_j^2=1$,
and Eq.~\eqref{eq:cofactor} gives $T=1/(1+E^{2n})$.

\textit{Integrals.} A contour integral over the sector of opening angle
$\pi/n$, which encloses only the pole at $e^{i\theta}$, gives
$\int_0^\infty dx/(1+x^{2n})=\pi/(2n\sin\theta)$, so
$I_1=1/(2n\sin\theta)=\Gamma/n$. Since $ET'=-2nT(1-T)$ and $ET\to0$ at both
ends, integration by parts gives $I_2=I_1/(2n)$.

For $n=2$, $\chi_2=E^2+i\sqrt2E-1$, $t_1=\Gamma=1/\sqrt2$ and
$I_1=1/(2\sqrt2)$; for $n=3$, $\chi_3=(E+i)(E^2+iE-1)$, $I_1=1/3$ and
$I_2=1/18$. Since $F_0=1/(2n)$, Eq.~\eqref{eq:Pformula} gives
$P_{\rm tr}=Aq/n+\beta(A)$. Splitting the affinity at $\log n$ as in
Sec.~\ref{sec:lower}, and evaluating at $A=\log n$, gives
$\log n/(2n)\le\min_AP_{\rm tr}\le6\log n/n$ for $n\ge3$: the window edges of
this profile carry noise of order $1/n$ rather than $1/n^2$.

\section{Finite-dimensional calibration}
\label{sec:calibration}

This section checks that our counting conventions reproduce the
finite-dimensional values of Ref.~\cite{Menczel_2021tu}, and records a
four-level member below the reported four-level value. The models of
Ref.~\cite{Menczel_2021tu} have one channel, $\gamma=1$, a ladder jump
$L=\sum_kl_k|k\rangle\langle k+1|$, and a total Hamiltonian with zero
diagonal. We write $\bar n$ for the occupation of their single channel. With
$H_0=\operatorname{diag}(0,1,\dots,N-1)$, $A=\log[(1+\bar n)/\bar n]$, $T=1/A$,
and $X=H-H_0$, every such model belongs to the class with $[H_0,L]=-L$, and
with one channel $\sigma=Aj$ and $P=Ad/j$.

\textit{Replay of the printed parameters.} Table~\ref{tab:calibration} lists
the printed parameters. Evaluating Eq.~\eqref{eq:poisson} at these values,
taken as exact, gives the products in the table. Two independent
implementations, one solving the stationary and response equations and one
differentiating the tilted eigenvalue numerically at 50-digit precision,
agree to $3\times10^{-15}$. For $N=2$, $j=1/4$ and
$P=A[(2\bar n+1)-3/(4(2\bar n+1))]$ exactly at $H_{12}=(2\bar
n+1)/(2\sqrt2)$. The values for $N=2$ and $N=3$ agree with
Ref.~\cite{Menczel_2021tu} to the printed digits. The four-level value is
$0.262667$, against the reported $0.2625$.

\begin{table*}
\caption{Printed parameters of Ref.~\cite{Menczel_2021tu} (upper triangle of
$H$; the lower triangle is conjugate) and the products they give in our
conventions, compared with the reported values.}
\label{tab:calibration}
\begin{ruledtabular}
\begin{tabular}{cllll}
$N$ & $\bar n$; $l_k$ & $H$ & $P$ (this work) & reported\\
\hline
2 & $0.02779$; $1$ & $H_{12}=(2\bar n+1)/(2\sqrt2)$ & $1.245872$ & $1.2459$\\
3 & $0.01096$; $1$, $0.03357$ & $H_{12}=0.1681$, $H_{23}=0.3347$, $H_{13}=0.2164i$ & $0.472421$ & $0.47242$\\
4 & $0.00560$; $1$, $0.0280$, $0$ & $H_{12}=0.101$, $H_{23}=0.343-0.025i$, $H_{34}=-0.206-0.012i$, & $0.262667$ & $0.2625$\\
 & & $H_{13}=0.008+0.132i$, $H_{24}=-0.002-0.059i$, $H_{14}=0.139$ & & \\
\end{tabular}
\end{ruledtabular}
\end{table*}

\subsection{Exact enclosures of the four-level values}
\label{sec:fourlevel}

Whether the difference from the reported four-level value is ours, and
whether a four-level member does better, is settled by a computer-assisted
evaluation in exact arithmetic. It shows three things. Part (i): at the printed
four-level parameters, $P>0.26255$. Part (ii): the products at
$\operatorname{Re}H_{23}=0.34251$ and at $0.34349$, both of which round to the
printed $0.343$, differ by more than $10^{-4}$. Part (iii): the four-level member
with $\bar n=0.005399090$, $l=(1,0.030296333,0.004094694)$, zero diagonal,
and $H_{12}=0.099280080$, $H_{23}=0.348078039+0.000002033i$,
$H_{34}=-0.208739471+0.000015192i$, $H_{13}=-0.000000614+0.132090823i$,
$H_{24}=-0.000004446-0.067632678i$, $H_{14}=0.147669178-0.000010122i$ (all
decimals exact) has a unique, faithful, attracting stationary state and
$P=0.2611508361761\ldots$; hence $\Pmin^4<0.26116$.

\textit{Exact arithmetic.} Each case is one $16\times16$ matrix $\mathsf M$ representing
the generator on $4\times4$ matrices, with rational complex entries. The test
applied is exact. Replacing the first row of $\mathsf M$ by the trace
functional gives a matrix $\mathsf M'$, inverted in exact rational
arithmetic. The stationary state $\rho$, the image under $\mathsf M'^{-1}$ of
the first unit vector, and the response $\eta$ of Eq.~\eqref{eq:poisson} are
then exact rationals, verified by substitution with zero residual, and $j$ and
$d$ are exact positive rationals. In all four cases the leading principal
minors of the exact $\rho$ are positive, so $\rho\succ0$. The only irrational
quantity is $A=\log[(1+\bar n)/\bar n]$. We write $(1+\bar n)/\bar n=2^kx$
with $1\le x<2$, so $A=k\log2+\log x$; in (i) and (ii),
$A=7\log2+\log(1257/896)$. Each logarithm is enclosed between rational bounds
by the series $\log x=2\sum_{j\ge0}z^{2j+1}/(2j+1)$, $z=(x-1)/(x+1)\le1/3$,
truncated after $\nu$ terms. The positive tail is at most
$2z^{2\nu+1}/[(2\nu+1)(1-z^2)]$. We take $\nu=40$ in (i) and (ii), and
$\nu=80$ in (iii). Multiplying by the exact $d/j$ encloses $P$ in a rational
interval narrower than $10^{-40}$ in (i) and (ii) and $10^{-79}$ in (iii).
These intervals are decisive because they are rigorous and far narrower than
the margins. The lower end in (i) exceeds $0.26255$, the gap between the two
intervals in (ii) exceeds $10^{-4}$, and the upper end in (iii) is below
$0.26116$. The cases are finite in number and each is evaluated completely.
In (iii), $L^\dagger L=\operatorname{diag}(0,1,l_2^2,l_3^2)$ has distinct
entries, so a matrix commuting with $L$ and $L^\dagger$ is diagonal and then
scalar. In (i) and (ii), $l_3=0$, but the pairs of diagonal entries of
$L^\dagger L$ and $LL^\dagger$ still separate the four levels. A commuting
matrix is therefore diagonal, and commuting with $L$ makes it
$a\Pi_{123}+b\Pi_4$, with $\Pi_{123}$ and $\Pi_4$ the projections on levels
$1$--$3$ and on level $4$. In Step 3 of Sec.~\ref{sec:relaxation},
$[H,Y]=\xi Y$ for such a $Y$ gives, in the entry $(1,4)$, $H_{14}(b-a)=0$;
since $H_{14}=0.139\neq0$, $Y$ is scalar and $\xi=0$. The argument of Steps 3
and 4, with this commutant in place of Step 1, then applies in all four cases.
Zero is algebraically simple, every other eigenvalue has negative real part,
and every density matrix relaxes to $\rho$, so Eq.~\eqref{eq:poisson} gives
the long-time cumulants. For (iii), a separate evaluation of the tilted
eigenvalue in 60-digit arithmetic agrees with the exact cumulants to
$10^{-17}$.

Part (i) shows that the gap between $0.262667$ and the reported $0.2625$ is
not a rounding error of our computation. Part (ii) shows that the printed
parameters cannot fix the fourth digit. Inputs indistinguishable at the
printed precision give products differing by more than twice the half-width
$5\times10^{-5}$ of the reported last digit. The cause of the difference is
therefore not decidable from the published table; both values agree with the
summary value $0.26$ of Ref.~\cite{Menczel_2021tu}. Part (iii) shows that the
four-level ladder model of Ref.~\cite{Menczel_2021tu} reaches below its
reported optimum, so $\Pmin^4<0.26116<0.2625$. These values calibrate the
conventions; none of them enters the proofs of the main text.

\end{document}